\documentclass[10pt,journal,twoside,final]{IEEEtran}

\IEEEoverridecommandlockouts

\makeatletter
\newcommand{\biggg}[1]{{\hbox{$\left#1\vbox to 20.5pt{}\right.\n@space$}}}
\newcommand{\Biggg}[1]{{\hbox{$\left#1\vbox to 23.5pt{}\right.\n@space$}}}
\newcommand{\bigggg}[1]{{\hbox{$\left#1\vbox to 26.5pt{}\right.\n@space$}}}
\newcommand{\Bigggg}[1]{{\hbox{$\left#1\vbox to 29.5pt{}\right.\n@space$}}}
\newcommand{\biggggg}[1]{{\hbox{$\left#1\vbox to 32.5pt{}\right.\n@space$}}}
\newcommand{\Biggggg}[1]{{\hbox{$\left#1\vbox to 35.5pt{}\right.\n@space$}}}
\newcommand{\bigggggg}[1]{{\hbox{$\left#1\vbox to 38.5pt{}\right.\n@space$}}}
\newcommand{\Bigggggg}[1]{{\hbox{$\left#1\vbox to 41.5pt{}\right.\n@space$}}}
\makeatother

\usepackage{amsfonts}
\usepackage{bm,cite,float,amssymb}
\usepackage{subfigure,amsfonts}
\usepackage{mdwmath}
\usepackage{mdwtab}
\usepackage{xcolor,color}
\usepackage{cool}
\usepackage{graphicx}
\usepackage{multicol}
\usepackage{amsmath,lipsum}
\usepackage{balance}
\usepackage{diagbox}
\usepackage{bm}
\usepackage{cuted}
\usepackage{stfloats}
\usepackage{algorithm}%
\usepackage{algorithmicx}
\usepackage{multirow}%
\usepackage{amsthm}%
\usepackage{textcomp}%
\usepackage{manyfoot}%
\usepackage{booktabs}%
\usepackage{algpseudocode}

\graphicspath{{imgs/}}   

\makeatletter
\renewcommand\paragraph{\@startsection{paragraph}{4}{\z@}%
            {-2.5ex\@plus -1ex \@minus -.25ex}%
            {1.25ex \@plus .25ex}%
            {\normalfont\normalsize\itshape}}
\makeatother
\usepackage{epstopdf}
\usepackage[framemethod=0]{mdframed}
\usepackage{showexpl}

\usepackage{cite}

\mdfdefinestyle{exampledefault}{%
rightline=true,innerleftmargin=10,innerrightmargin=10,
frametitlerule=true,frametitlerulecolor=green,
frametitlebackgroundcolor=yellow,
frametitlerulewidth=2pt}

\allowdisplaybreaks

\definecolor{myGreen}{rgb}{0.76, 0.93, 0.63}
\definecolor{myGreen2}{rgb}{0.92, 0.92, 0.92}

\begin{document}

\title{{Maneuverable-Jamming-Aided Secure Communication and Sensing in A2G-ISAC Systems}}

\author{Libiao Lou, 
Yuan Liu, 
Fotis Foukalas, 
Hongjiang Lei, 
Gaofeng Pan,  
Theodoros A. Tsiftsis,  
and Hongwu~Liu
\thanks{The work of T. A. Tsiftsis was supported by the Project NEURONAS. The research project NEURONAS is implemented in the framework of H.F.R.I call ``3rd Call for H.F.R.I.’s Research Projects to Support Faculty Members \& Researchers'' (H.F.R.I. Project Number: 25726).}
\thanks{L. Lou, Y. Liu, and H. Liu are with the School of Information Science and Electrical Engineering, Shandong Jiaotong University, Jinan 250357, China (e-mail: 23208018@sdjtu.edu.cn, 240041@sdjtu.edu.cn,  liuhongwu@sdjtu.edu.cn).}
\thanks{F. Foukalas is with the Department of Informatics and Telecommunications, University of Thessaly, 35100 Lamia, Greece (e-mail: foukalas@ieee.org).}
\thanks{H. Lei is with the School of Communications and Information Engineering, Chongqing University of Posts and Telecommunications, Chongqing 400065, China (e-mail: leihj@cqupt.edu.cn).}
\thanks{G. Pan is with  the School of Cyberspace Science and Tech
nology, Beijing Institute of Technology, Beijing 100081, China (e-mail:
 gaofeng.pan.cn@ieee.org).}
 \thanks{T. A. Tsiftsis is with the Department of Informatics and Telecommunications, University of Thessaly, Lamia 35100, Greece, and also with the Department of Electrical and Electronic Engineering, University of Nottingham Ningbo China, Ningbo 315100, China (e-mail: tsiftsis@uth.gr).}
}
 
\maketitle
\setcounter{page}{1}
\begin{abstract}
\textcolor{black}{In this paper, we propose a maneuverable-jamming-aided secure communication and sensing (SCS) scheme for an air-to-ground integrated sensing and communication (A2G-ISAC) system, where a dual-functional source UAV and a maneuverable jamming UAV operate collaboratively in a hybrid monostatic-bistatic radar configuration. The maneuverable jamming UAV emits artificial noise to assist the source UAV in detecting multiple ground targets while interfering with an eavesdropper. The effects of residual interference caused by imperfect successive interference cancellation on the received signal-to-interference-plus-noise ratio are considered, which degrades the system performance. To maximize the average secrecy rate (ASR) under transmit power budget, UAV maneuvering constraints, and sensing requirements, the dual-UAV trajectory and beamforming are jointly optimized. Given that secure communication and sensing fundamentally conflict in terms of resource allocation, making it difficult to achieve optimal performance for both simultaneously, we adopt a two-phase design to address this challenge. By dividing the mission into the secure communication (SC) phase and the SCS phase, the A2G-ISAC system can focus on optimizing distinct objectives separately. 
In the SC phase, a block coordinate descent algorithm employing the trust-region successive convex approximation and semidefinite relaxation iteratively optimizes dual-UAV trajectory and beamforming. For the SCS phase, a weighted distance minimization problem determines the suitable dual-UAV
sensing positions by a greedy algorithm, followed by the joint optimization of source beamforming and jamming beamforming. Simulation results demonstrate that the proposed scheme achieves the highest ASR among benchmarks while maintaining robust sensing performance, and confirm the impact of the SIC residual interference on both secure communication and sensing.}

\end{abstract}


\section{Introduction} 

Integration of sensing and communication (ISAC) is essential for sixth generation wireless networks to enable extremely ultra-reliable and extremely low-latency connectivity while supporting real-time environmental perception and data transmission\cite{ISAC_Dual-Functional_Wireless_Networks,ISAC_Resource_Allocation,CS_MIMO_ISAC}. By unifying both communication and radar sensing functionalities, ISAC improves spectral efficiency, realizes environmental awareness, and paves the way for diverse intelligent applications like autonomous systems, immersive digital-physical interactions, and smart city infrastructures \cite{ISAC_Advances_Challenges}. \textcolor{black}{Unmanned aerial vehicles (UAVs) serve as transformative enablers for air-to-ground (A2G)-ISAC systems, providing dynamic network deployment, high-resolution sensing, and reliable line-of-sight (LoS) transmissions \cite{UAV_ISAC_EE, beamforming_design, UAV_ISAC_Deployment_and_Precoder}. UAV-aided A2G-ISAC enables diverse applications including disaster response, precision agriculture, smart city monitoring, and autonomous vehicle navigation by leveraging UAV maneuverability and communication-sensing synergy \cite{UAV_ISAC_Sky, UAV_ISAC_IoT}.}

\textcolor{black}{Due to the open nature of wireless communications, LoS propagation vulnerabilities expose A2G-ISAC systems to security threats including eavesdropping and spoofing. Physical layer security (PLS) technology provides crucial protection with minimal computational requirements. In dynamic A2G-ISAC environments, PLS techniques such as beamforming, artificial noise (AN) jamming, trajectory design, and resource allocation effectively mitigate eavesdropping and ensure secure wireless transmissions}\textcolor{black}{\cite{UAV_ISAC_Security_Aware,Secrecy_AAV_ISAC,secure_UAV_ISAC_JBT,UAV_ISAC_Coverage_and_Security,UAVs_meet_ISAC_Secure,Security_ISAC_IRS_UAV,Secure_IRS_UAV_ISAC}.} \textcolor{black}{ Yang $et$ $al.$ \cite{UAV_ISAC_Security_Aware} employed a UAV-enabled ISAC platform to communicate with terrestrial users while sensing targets in pre-determined areas. In \cite{Secrecy_AAV_ISAC} and \cite{secure_UAV_ISAC_JBT}, a joint UAV trajectory and beamforming optimization method was proposed to maximize the average secrecy rate (ASR) while ensuring sensing accuracy for untrusted targets in A2G-ISAC systems. Furthermore, Benaya $et$ $al.$ \cite{UAV_ISAC_Coverage_and_Security} considered the deployment of a UAV to transmit jamming signals that disrupt communication links with malicious targets or potential eavesdroppers identified through sensing.} To achieve the high-precision tracking and maximize the secrecy rate, a real-time UAV trajectory optimization method utilizing the extended Kalman filtering and coordinated beamforming was proposed in \cite{UAVs_meet_ISAC_Secure}. With the advancements of intelligent reflecting surface (IRS), several IRS-aided PLS schemes were proposed for A2G-ISAC systems to enhance both the security and sensing performance by jointly optimizing the UAV trajectory and beamforming \cite{Security_ISAC_IRS_UAV,Secure_IRS_UAV_ISAC}.

Nevertheless, most research on the PLS-based secure communication for A2G-ISAC systems remains confined to single-UAV scenarios\textcolor{black}{\cite{UAVs_meet_ISAC_Secure,Security_ISAC_IRS_UAV,Secure_IRS_UAV_ISAC}}, failing to fully exploit the performance gains enabled by multi-UAV coordination. Specifically, existing approaches predominantly assume static or predefined flight trajectories, overlooking the inherent spatial diversity and cooperative beamforming capabilities offered by distributed UAVs\textcolor{black}{\cite{Dual_UAV_secure,Multi_UAV_Ssecure,LAE_ISAC_BEAM_TRAJECTORY}}. 
Through coverage extension, interference suppression, and resource scheduling, 
multi-UAV coordination effectively improved system gains for A2G-ISAC\textcolor{black}{\cite{Multi_UAV_ISAC_GCWkshps, Multi_UAV_ISAC_WCSP, Multi_UAV_ISAC_IoTJ,EKF_multi_UAV_JBT}}. In an A2G-ISAC system consisting of multiple dual-functional UAVs, the collaborative signal processing, UAVs' placement, and power control were jointly explored to enhance the detection performance of the ground target \cite{Multi_UAV_ISAC_GCWkshps}. In addition, the beamforming and UAVs' placement were considered to maximize the minimum detection probability over the target, while ensuring the minimum signal-to-interference-plus-noise ratio (SINR) requirement among ground users \cite{Multi_UAV_ISAC_WCSP}. When multiple dual-functional UAVs were deployed to communicate with multiple ground users and detect a ground target, the user association, beamforming, and UAV trajectory were jointly designed to maximize the weighted sum-rate,  while ensuring the sensing beampattern gain of the target \cite{Multi_UAV_ISAC_IoTJ}. \textcolor{black}{In addition, the evolution of A2G-ISAC systems is characterized by a clear shift from static optimization toward dynamic real-time processing. In \cite{EKF_multi_UAV_JBT}, real-time estimation of mobile user positions is directly utilized for beamforming design and trajectory optimization.} \textcolor{black}{Although multi-UAV coordination can improve communication and sensing resource allocation, performing simultaneous optimization throughout the entire task time may cause resource competition. In contrast, Chai $et$ $al.$ \cite{Time_slot} proposed an approach dividing task time into two phases, where the first phase executes communication tasks, and the second phase optimizes sensing performance while maintaining communication capabilities. Nevertheless, this approach fails to meet secure communication requirements, particularly in countering eavesdropping threats.}

On the other hand, jamming techniques, including friendly and malicious jamming, were strategically employed in A2G-ISAC systems to degrade the eavesdropping reception quality or legitimate communication integrity, respectively\textcolor{black}{\cite{UAV_ISAC_Coverage_and_Security,UAV_ISAC_Malicious_Jamming,UAV_ISAC_Jamming_Attack,add_RIS_A2G_network,add_secure_Ad_Hoc,UAV_ISAC_PLS_Multiple_Eavesdroppers,UAV_secure_JTR,Jamming_multi_UAV_ISAC_secure}}. To mitigate adversarial impacts on the received SINRs, both friendly and malicious jamming necessitate robust beamforming and power allocation schemes in dynamic A2G-ISAC environments. 
By reducing the received SINRs for authorized users, malicious jamming disrupts legitimate communication links, thus degrading PLS performance in A2G-ISAC system {\cite{UAV_ISAC_Malicious_Jamming,UAV_ISAC_Jamming_Attack}}. \textcolor{black}{While Sun $et$ $al.$ \cite{add_RIS_A2G_network} introduced semantic communication for anti-jamming, their work overlooked ISAC waveforms. Similarly, Lin $et$ $al.$ \cite{add_secure_Ad_Hoc} proposed a terrestrial cooperative defense framework, whose effectiveness is limited by the lack of consideration for aerial mobility.}
Conversely, friendly jamming enhanced the PLS performance of A2G-ISAC systems by intentionally introducing controlled interference to obscure eavesdroppers, thereby improving secrecy rate and reducing information leakage risks \textcolor{black}{\cite{UAV_ISAC_Coverage_and_Security,UAV_ISAC_PLS_Multiple_Eavesdroppers,UAV_secure_JTR,Jamming_multi_UAV_ISAC_secure}}. In \cite{UAV_ISAC_PLS_Multiple_Eavesdroppers}, a stationary jamming UAV is deployed at the geometric center of multiple eavesdroppers to interfere  eavesdropping links, addressing PLS challenges in A2G-ISAC scenarios. \textcolor{black}{Subsequently, in \cite{UAV_ISAC_Coverage_and_Security} and \cite{UAV_secure_JTR}, the maneuverable jamming UAV first located malicious targets based on sensing results from the dual-functional base-station (BS) and then transmitted jamming signals to minimize the eavesdropper's SINR, the maximizing the secrecy rate.}  \textcolor{black}{Moreover, Wei $et$ $al.$ \cite{Jamming_multi_UAV_ISAC_secure} utilized the UAV to recycle AN to achieve dual-functional jamming and sensing, combining Kalman filtering and neural network predictor to adaptively adjust flight trajectory and resource allocation, thereby enhancing the dynamic deterrence capability against mobile eavesdroppers.} However, current research on cooperative jamming in A2G-ISAC systems lacks comprehensive consideration of jamming UAV mobility in PLS designs. This oversight restricts the potential for further enhancing PLS performance through dynamic trajectory optimization of jamming UAVs. 

\textcolor{black}{Motivated by the aforementioned works, we propose a maneuverable-jamming (MJ)-aided secure communication and sensing (SCS) scheme for an A2G-ISAC system. In the proposed scheme, a jamming UAV is deployed to simultaneously suppress eavesdropping links and enhance sensing performance through AN transmission, effectively constructing a hybrid monostatic-bistatic radar system with the source UAV. By jointly optimizing dual-UAV trajectory and beamforming, the SCS scheme achieves substantial performance gains in both secure communication and target sensing.} Our main contributions can be summarized as follows:

\begin{itemize}
\item \textcolor{black}{We propose an MJ-aided SCS scheme for A2G-ISAC system, where source and jamming UAVs operate within a hybrid monostatic-bistatic radar system. The maneuverable jamming UAV provides spatial degrees of freedom, which simultaneously enhance secure communication and improve performance via a secure inter-UAV control link. Specifically, the AN signals transmitted by the jamming UAV are designed to degrade eavesdropper reception and enhance target sensing capabilities concurrently. Considering imperfect successive interference cancellation (SIC) at the communication and radar receivers, residual interference effects on the received SINRs are modeled.} 
\item \textcolor{black}{To balance the inherent resource allocation trade-off between secure communication and sensing performance, we develop a two-phase optimization design that decouples the optimization process into sequential secure communication (SC) and SCS phases. For the complex non-convex ASR maximization problem, under constraints of transmit power budgets, UAV maneuvering, and sensing requirements, we propose a block coordinate descent (BCD) approach to jointly optimize the decoupled dual-UAV trajectory and beamforming subproblems, utilizing trust-region successive convex approximation (SCA) and semidefinite relaxation (SDR).}
\item \textcolor{black}{To determine appropriate dual-UAV sensing positions, we first optimize trajectories solely for the SC phase. Then, we propose a greedy algorithm to solve a corresponding weighted distance minimization problem for the SCS phase. Simulation results indicate the proposed scheme outperforms existing benchmarks in both secure communication and target sensing. The impact of residual interference on the A2G-ISAC system performance is further examined.}

\end{itemize}

The remainder of the paper is organized as follows. Section II establishes the system model of the considered A2G-ISAC system and formulates the ASR maximization problem. Sections III and IV present the suboptimal solutions for the SC and SCS purposes, respectively. \textcolor{black}{Section V clarifies the system performance through simulations, while Section VI concludes the work.}
 
$Notations$: The statistical expectation is denoted as $\mathbb{E}(\cdot)$. Vectors and matrices are denoted using bold lowercase and uppercase italic letters, respectively. $(\cdot)^*$, $(\cdot)^T$, and $(\cdot)^H$ denote the conjugate, transpose, and Hermitian transpose operations, respectively.  $\bm{0}$ and $\bm{I}$ stand for the all-zero matrix and identity matrix, respectively. $\bm{A} \succeq 0$ represents that 
matrix $\bm A$ is positive semidefinite. $[\bm{A}]_{x,y}$ denotes the element in the $x$-th row and $y$-th column of  matrix $\bm A$. The magnitude and phase of $[\bm{A}]_{x,y}$ are denoted by $\big\vert[{\bm A}]_{x,y}\big\vert$ 
and $\theta^{\bm A}_{x,y}$, respectively. ${\rm{rank}}(\cdot)$ and ${\rm{tr}}(\cdot)$ denote the rank and trace of a matrix, respectively. The Euclidean norm, nuclear norm, and spectral norm are denoted by
$\| \cdot \|$, $\| \cdot \|_*$, and $\| \cdot \|_2$, respectively.  $\mathbb{C}^{M \times N}$ denotes the $M \times N$ complex matrices. $\mathcal{C}\mathcal{N}(\nu, \sigma^2)$ is the circularly symmetric complex Gaussian distribution with mean $\nu$ and variance $\sigma^2$.

\section{System Model}

\begin{figure}[t]
 \begin{center}
    \includegraphics[width=3.4in,  height=1.8in]{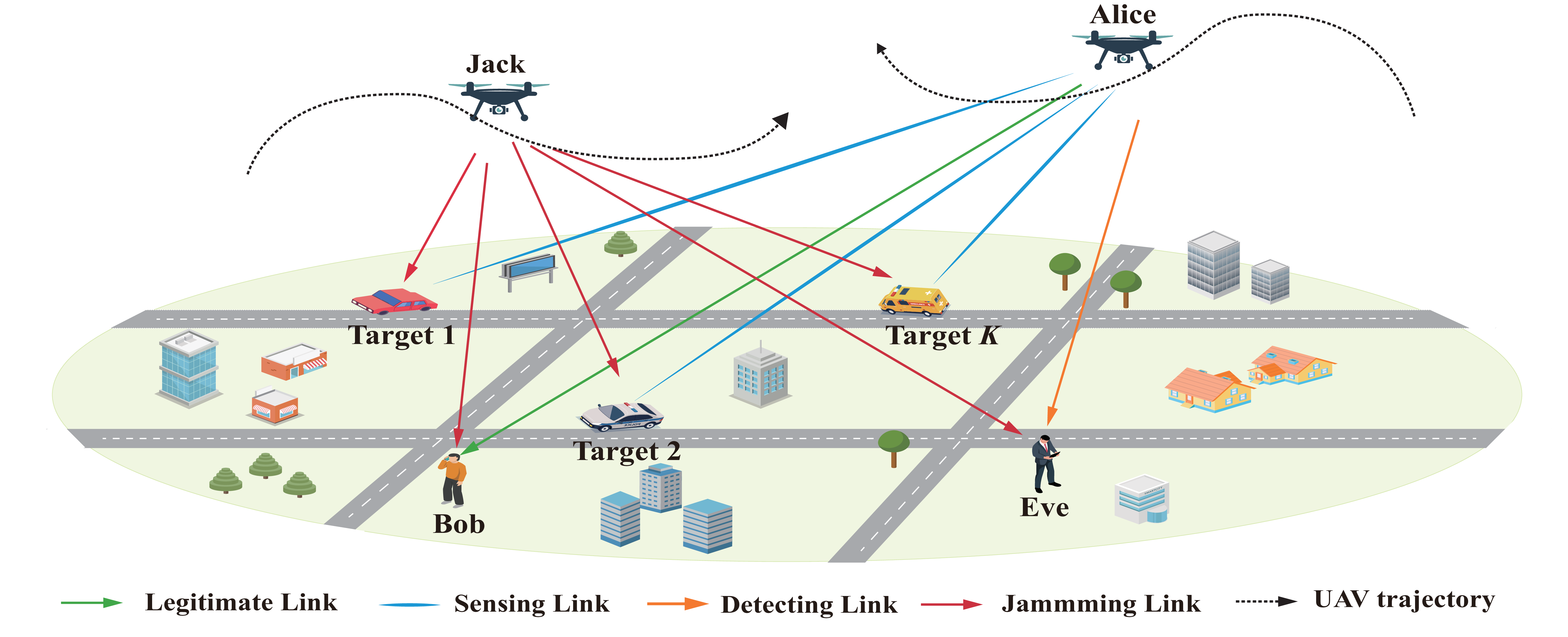}
    \caption{\textcolor{black}{MJ-aided A2G-ISAC system model.}}
    \label{fig:system_model}
\end{center}
\vspace{-0.3in}
\end{figure}

\textcolor{black}{The considered A2G-ISAC system, as shown in Fig. \ref{fig:system_model}, includes a source UAV (Alice), a jamming UAV (Jack), a legitimate ground user (Bob), an eavesdropper (Eve), and $K$ ground targets. Alice functions as a dual-functional aerial BS, transmitting confidential signals to Bob while performing radar sensing of the $K$ targets \cite{ISAC_Dual-Functional_Wireless_Networks,ISAC_Resource_Allocation}. This modeling approach enables us to concentrate on the core challenges of MJ-aided SCS scheme design, without introducing other interfering factors. To enhance both secure communication and sensing, Jack transmits AN signals to degrade Eve’s reception and assist in target detection. The target echoes from Alice’s ISAC waveforms and Jack’s AN signals are combined at Alice under a hybrid monostatic-bistatic radar system \cite{Hybrid_Bistatic_Monostatic,Hybrid_Multistatic}. In addition, each UAV is equipped with a uniform linear array consisting of $M$ antenna elements, while Bob and Eve each have a single receive antenna.}

\textcolor{black}{To enable secure transmission and target sensing, the dual-UAV system operates during a finite task period $\mathcal{T} \triangleq [0,T]$ divided into $N$ time slots of duration $\Delta_t = T/N$. The slot duration $\Delta_t$ is minimized to ensure constant dual-UAV locations within each slot, facilitating joint dual-UAV trajectory and beamforming design. The discrete time slot model ensures synchronization across all nodes, while the quasi-static channel assumption mitigates the need for rapid resynchronization \cite{Multi_UAV_ISAC_IoTJ}. In time slot $n \in \mathcal{N} \triangleq \{1, 2, \ldots, N\}$, UAV $m$ is located at $[x_m[n], y_m[n], H_m]$, where $m\in\{a,j\}$ represents Alice and Jack, respectively, and $H_m$ is the fixed altitude to simplify problem formulation while maintaining extensibility for three-dimensional trajectory optimization \cite{beamforming_design,UAVs_meet_ISAC_Secure}.} Moreover, the horizontal location of UAV $m$ is denoted by ${\bm{u}}_{m}[n]= \big[{x}_{m}[n],{y}_{m}[n] \big]^T$. The location of ground node $q$ is represented by ${\bm{v}}_{q}=[{x}_{q},{y}_{q}]^T$, where $q \in \{b, e\}$ represents Bob and Eve, respectively. The initial and final horizontal locations of UAV $m$ are respectively defined as ${\bm u}_m^{_{\rm I}} = [x_m^{{\rm I}}, y_m^{_{\rm I}}]^T$ and ${\bm u}_m^{_{\rm F}} = [x_m^{_{\rm F}}, y_m^{_{\rm F}}]^T$. \textcolor{black}{Let $D_{\max} = {V}_{\max} \Delta_t$ denote the maximum UAV displacement in each time slot, where ${V}_{\max}$ denotes the maximum flight speed of each UAV, and $d_{\min}$ is the minimum safety separation distance between two UAVs.} During a task period, the locations of the dual-UAV are limited by the following UAV maneuvering constraints

\begin{eqnarray}
  & ~~~~~~~~ {\bm u}_m[0] = {\bm u}_m^{_{\rm I}},~  {\bm u}_m[N] =  {\bm u}_m^{_{\rm F}},  \label{position_I_F} \\
& \Vert \bm{u}_m[n]-\bm{u}_m[n-1]\Vert \le D_{\max}~, ~\forall n \in\mathcal{N} , \label{displacement}\\ 
 &\textcolor{black}{\Vert \bm{u}_a[n]-\bm{u}_j[n]\Vert^2 +(H_a-H_j)^2 \ge d_{\min}^2~, ~\forall n \in\mathcal{N} .}
\label{collision}
\end{eqnarray}
Within a task period, the total time slots are classified into two phases, namely, the SC and SCS phases, which are denoted as $\mathcal{N}_c \triangleq \{1,2\cdots,N_c\}$ and $\mathcal{N}_s \triangleq \{1,2\cdots,N_s\}$, respectively, where $N_c$ + $N_s$ = $N$. \textcolor{black}{This two-phase design sequentially prioritizes mission objectives, first establishing a secure communication link, and then activating sensing and communication functions \cite{Time_slot}.} \textcolor{black}{Specifically, in the SC phase, Alice and Jack collaborate to enhance secure communication, while in the SCS phase, they conduct both secure communication and target sensing cooperatively.}
In time slot $n$, the signals transmitted by Alice and Jack can be respectively expressed as
\begin{equation}
    {\bm{x}}_a[n] = \left\{ {\begin{array}{*{20}{c}}
{ {\bm{w}_{a}}[n]s_{a}[n] },& {\forall n \in {\cal N}_c},\\
{ {\bm{w}_{a}}[n]s_{a}[n]+ \bm{s}_r[n] },&{\forall n \in {\cal N}_s},
\end{array}} \right.
 \label{eq:transmit_sensing_signal}    
\end{equation}
and
\begin{eqnarray}
 {\bm{x}}_{j}[n]={\bm{w}}_{j}[n]{s}_{j}[n] , \qquad   \forall \mathit{n} \in \mathcal{N},
 \label{eq:transmit_AN}
\end{eqnarray}
where ${\bm{w}}_{a}[n]\in {\mathbb{C}}^{M\times1}$ and ${\bm{w}}_{j}[n]\in {\mathbb{C}}^{M\times1}$ are the transmit beamformers at Alice and Jack, respectively, ${s}_{a}[n] \backsim \mathcal{C}\mathcal{N}(0,1)$ is Alice's information signal intended to Bob, ${s}_{j}[n]\backsim \mathcal{C}\mathcal{N}(0,1)$ is the AN signal transmitted by Jack, and ${\bm{s}}_{r}[n]\in {\mathbb{C}}^{M\times1}$ is the sensing signal with zero mean and covariance matrix ${\bm{R}}_{r}[n]=\mathbb{E}( {\bm{s}}_{r}[n]{\bm{s}}_{r}^{H}[n] )\succeq 0$. We assume that ${s}_{a}[n]$ and ${\bm{s}}_{r}[n]$ are independent of each other.  
Based on \eqref{eq:transmit_sensing_signal} and \eqref{eq:transmit_AN}, ${s}_{a}[n]$ and ${s}_{j}[n]$ are respectively transmitted using a single beam,  while ${\bm{s}}_{r}[n]$ is transmitted using ${M}_{s}$ beams, where $0\leq {M}_{s}\leq M$. Specifically, the ${M}_{s}$ beams are generated by performing eigenvalue decomposition on ${\bm{R}}_{r}[n]$ with ${\rm{rank}}(\bm{R}_r[n])=M_s$. The transmit powers of Alice and Jack satisfy 
\begin{equation}
\mathbb{E}(\|\bm{x}_a[n]\|^2) \!=\! \left\{ {\begin{array}{*{20}{c}}
\Vert\bm{w}_a[n]\Vert^2\le P_a^{\rm{max}},~   n\in\mathcal{N}_c,\\
\Vert\bm{w}_a[n]\Vert^2+ {\rm{tr}}(\bm{R}_r[n])\le P_a^{\rm{max}}, ~  n\in\mathcal{N}_s,
\end{array}} \right.
 \label{power_alice}    
\end{equation}
and
\begin{eqnarray}
\mathbb{E}(\|\bm{x}_j[n]\|^2)=\Vert\bm{w}_j[n]\Vert^2\le P_j^{\rm{max}},~   n\in\mathcal{N} ,
 \label{power_jack}
\end{eqnarray}
 where $\mathbb{E}(\|\bm{x}_m[n]\|^2)$ and $P_m^{\max}$ denote the average transmit power and the maximum transmit power of UAV $m$.

\subsection{Secure Communication Model}

Considering that minimum maneuvering altitude of each UAV is relatively high, 
in general strong LoS links exist between UAVs and ground nodes  \textcolor{black}{\cite{UAV_ISAC_Deployment_and_Precoder,UAV_ISAC_Sky, UAV_ISAC_IoT}}. Therefore, 
we adopt the LoS channel model in this work and the channel from UAV $m$ to ground node $q$ is represented as  
\begin{eqnarray}
{\bm{h}}_{mq}[n]&\!\!\!\!\!=\!\!\!\!\!&\bm{a}_{mq}[n]\sqrt{\beta {d}^{-2}_{mq}[n]}\nonumber \\ 
&\!\!\!\!\!=\!\!\!\!\!&\bm{a}_{mq}[n]\sqrt{\frac{\beta }{{\|{\bm{u}}_{m}[n]-{\bm{v}}_{q}\|}^{2}+{H}_{m}^{2}}}, 
\label{communication_channel}
\end{eqnarray}
where $\beta$ denotes the path-loss at the reference distance of ${d}_{0} = 1$ m and ${d}_{mq}[n]=\sqrt{{{\|\bm{u}_{m}[n]-\bm{v}_{q}\|}^{2}+{H}_{m}^{2}}}$ represents the distance between UAV $m$ and ground node $q$. \textcolor{black}{Following common practice in A2G-ISAC systems \cite{UAV_ISAC_Security_Aware,Secrecy_AAV_ISAC,secure_UAV_ISAC_JBT}, perfect channel state information is assumed to establish a tractable benchmark for evaluating the proposed MJ-aided SCS scheme.} In \eqref{communication_channel},  $\bm{a}_{mq}[n]$ is the steering vector, which is given by
\begin{eqnarray}
\bm{a}_{mq}[n]\!=\!{\!\left[ 1,{e}^{\mathfrak{j}2\pi \frac{{d}_{m}}{\lambda}\cos\theta_{mq}[n]} , \!\cdots \! ,{e}^{\mathfrak{j}2\pi \frac{{d}_{m}}{\lambda}(M-1)\cos\theta_{mq}[n] }\right]\!}^{T}\!\!\!,\!\!\!
\label{steering_vector_mq}
\end{eqnarray}
where $d_m$ and $\lambda$ denote the spacing between adjacent antennas and the wavelength of the carrier signal, respectively, and $\theta_{mq}[n]$ is the angle of departure from UAV $m$ to ground node $q$, which can be expressed as
\begin{eqnarray}
    \theta_{mq}[n]= \arccos \frac{{H}_{m}}{\sqrt{{{\|\bm{u}_{m}[n]-\bm{v}_{q}\|}^{2}+{H}_{m}^{2}}}}.
    \label{AoS_lq}
\end{eqnarray} 
In both the SC and SCS phases, the received signal at Bob in the time slot $n$ can be written as
\begin{eqnarray}
\!\!\!\!\!\!\!\!y^c_b[n] &\!\!\!\!=& \!\!\! \bm{h}_{ab}[n] \bm{x}_a [n]+\bm{h}_{jb}[n] \bm{x}_j[n]+ {z}_{b}[n] \nonumber \\
&\!\!\!\! =&  \!\!\!\bm{h}_{ab}[n]\bm{w}_a [n]s_a[n]+\bm{h}_{jb}[n]\bm{w}_j[n] s_j[n]+z_b[n]
\label{receive_signal_c}
\end{eqnarray}
and
\begin{eqnarray}
\!\!\!\!\!\!\!\!y^s_b[n] &\!\!\!\!=& \!\!\! \bm{h}_{ab}[n] \bm{x}_a [n]\!+\! \bm{h}_{jb}[n] \bm{x}_j[n]+ {z}_{b}[n] \nonumber \\
&\!\!\!\! =&  \!\!\!\bm{h}_{ab}[n](\bm{w}_a [n]s_a[n]\!+\!\bm{s}_r[n])\!+\!\bm{h}_{jb}[n]\bm{w}_j[n] s_j[n] \nonumber \\
&\!\!\!\!  &  \!\!\!  + z_b[n],
\label{receive_signal_s}
\end{eqnarray}
respectively, where ${z}_{b}\left [ n\right]\backsim \mathcal{C}\mathcal{N}( 0,{\sigma_b }^{2})$ is the additive white Gaussian noise at Bob. 

In the SC phase, the AN signal ${{s}}_j[n]$ is first detected and removed from the received signal by using SIC at Bob. Then, the intended information signal ${{s}}_a[n]$ is detected. \textcolor{black}{In the SCS phase, Bob sequentially decodes and cancels the sensing signal ${\bm{s}}_r[n]$ and the AN signal ${{s}}_j[n]$, which leaves residual interference proportional to $\varphi_{rb}\bm{h}_{ab}^H[n] \bm{R}_r[n] \bm{h}_{ab}[n]$ and $\varphi_{jb} |\bm{h}_{ab}^H[n] \bm{w}_j[n]|^2$, respectively. Subsequently, Bob decodes the desired information signal $s_a[n]$ from the remaining composite signal}.
\textcolor{black}{However, imperfect SIC introduces residual interference, which degrades system performance \cite{LAE_ISAC_BEAM_TRAJECTORY,CR_NOMA_SIC}.} To accurately model the residual interference effect, the received SINR at ground node $q$ in the SC and SCS phases are respectively expressed as
\begin{eqnarray}
{\gamma}^c_{q}[n] =   \frac{|\bm{h}_{aq}^H[n]\bm{w}_a[n]|^2}{ \varphi_{jq}|\bm{h}_{jq}^H[n]\bm{w}_j[n]|^2\!+\! \sigma_q^2}
\label{R^c_b}
\end{eqnarray}
and
\begin{eqnarray}    
   \gamma^s_q[n]=\frac{|\bm{h}_{aq}^H[n]\bm{w}_a[n]|^2}{\varphi_{rq}\bm{h}_{aq}^H[n]\bm{R}_r[n]\bm{h}_{aq}[n] + \varphi_{jq} |\bm{h}_{jq}^H[n]\bm{w}_j[n]|^2 + \sigma_q^2}, \!\!\!\!\!\!\!\!\!\!\!\!\nonumber \\
\!\!\!\!\!\!\!\!\!\ 
    \label{SINR_b} 
\end{eqnarray}
where $\varphi_{jq}$ ($0 \leq \varphi_{jq} \leq 1$) and $\varphi_{rq}$ ($0 \leq \varphi_{rq} \leq 1$) denote the residual interference levels after eliminating the AN signal and sensing signal, respectively. Then, for the considered A2G-ISAC system, the secrecy rates in the SC and SCS phases are respectively given by
\begin{eqnarray}
    {R}^c[n]=\big[\log_2(1 + \gamma^c_b[n])-\log_2(1 + \gamma^c_e[n])\big]^+
\label{Ach_rate_n_c}
\end{eqnarray}
and
\begin{eqnarray}
   {R}^s[n]=\big[\log_2(1 + \gamma^s_b[n])-\log_2(1 + \gamma^s_e[n])\big]^+,
\label{Ach_rate_n_s}
\end{eqnarray}
where $[x]^+ \triangleq {\max}(x,0)$ ensures the non-negative secrecy rate.

\subsection{Sensing Performance Metric}

\textcolor{black}{In the SCS phase, Alice and Jack collaboratively sense $K$ ground targets while maintaining secure communication. To improve the sensing performance, Alice and Jack construct a hybrid monostatic-bistatic radar system where Alice receives echoes containing both its own dual-function waveform and Jack's AN signal. The proposed scheme innovatively repurposes jamming signals as sensing resources, using Jack's AN as a known illuminator for target detection while degrading eavesdropper reception quality. By combining the high resolution of monostatic radar with the spatial diversity of bistatic radar, the system enhances target detection accuracy and coverage \cite{Hybrid_Bistatic_Monostatic,Hybrid_Multistatic}.}

\textcolor{black}{The $K$ ground targets are randomly located in an area of interest, where the location of target $k$ is denoted as $\bm{g}_k = [x_k, y_k]^T$, with $k \in \mathcal{K} \triangleq \{1, 2, \ldots, K\}$.} Similar to \eqref{communication_channel}, the channel from UAV $m$ to target $k$ can be expressed as
\begin{eqnarray}
\!\!\!\!\!\!\!\!\!{\bm{h}}_{mk}[n]&\!\!\!\!=\!\!\!\!&\bm{a}_{mk}[n] \sqrt{\beta {d}^{-2}_{mk}[n]}\nonumber \\
&\!\!\!\!=\!\!\!\!& \bm{a}_{mk}[n]\sqrt{\frac{\beta }{{\|{\bm{u}}_{m}[n]-{\bm{g}}_{k}\|}^{2}+{H}_{m}^{2}}}. 
\label{sensing_channel}
\end{eqnarray}
The incident sensing signal arrived at target $k$ is given by
\begin{eqnarray}
\!\!\!\!\!\!\!\!y_k^{s}[n] &\!\!\!\!\!=&\!\!\!\! \bm{h}_{ak}[n] \bm{x}_a [n]\!+\! \bm{h}_{jk}[n] \bm{x}_j[n] \nonumber \\
&\!\!\!\!\!=& \!\!\!\!\bm{h}_{ak}[n](\bm{w}_a [n]s_a[n]\!+\!\bm{s}_r[n])\!+\!\bm{h}_{jk}[n]\bm{w}_j[n] s_j[n].~
\end{eqnarray}
Then, the incident signal $y_k^{s}[n] $ is reflected back to Alice and Alice employs multi-angle signal fusion to detect target $k$ as performed in a hybrid monostatic-bistatic radar system.  
\textcolor{black}{Unlike conventional monostatic or bistatic radar systems, which contain a single transmitter, the proposed scheme leverages both source and jamming UAVs cooperatively. Specifically, Alice and Jack exchange state information via an inter-UAV communication link \cite{Multi_UAV_ISAC_WCSP,Multi_UAV_ISAC_IoTJ,EKF_multi_UAV_JBT}, with Alice serving as central coordinator to optimize trajectory and beamforming using combined data. Subsequently, Alice transmits the updated trajectory to Jack for local optimization, ensuring full utilization of both dual-functional waveforms and jamming signals for sensing.}
\textcolor{black}{Cram\'{e}r-Rao bound (CRB) is commonly adopted for radar sensing in ISAC systems \cite{ren2022fundamental}. However, the complexity of CRB in a hybrid monostatic-bistatic radar system, involving dual-UAV geometry, sensing SINR, and specific waveforms, makes it a challenge to jointly design the dual-UAV trajectory and beamforming \cite{Hybrid_Multistatic}. Therefore, our work instead employs a practical metric that incorporates beampattern gains from both Alice and Jack to quantify their detection contributions for target $k$.} Specifically, the distance-normalized sum-beampattern gain is 
introduced as

\begin{eqnarray}  
\zeta(\bm{u}_a[n], \bm{u}_j[n], \bm{g}_k) 
&\!\!\!\!\!=\!\!\!\!\!& \frac{\mathbb{E}\big[|\bm{a}^H_{ak}[n]\bm{x}_a[n]|^2\big]}{d^2_{ak}[n]} \!+\! \frac{\mathbb{E}\big[|\bm{a}^H_{jk}[n]\bm{x}_j[n]|^2\big]}{d^2_{jk}[n]}  \nonumber\\
&\!\!\!\!\! = \!\!\!\!\!& \frac{\bm{a}^H_{ak}[n](\bm{w}_{a}[n]\bm{w}_{a}[n]^H+
\bm{R}_r[n])\bm{a}_{ak}[n]}{ \Vert \bm{u}_a[n] - \bm{g}_k \Vert^2 + H_a^2}\nonumber\\
&\!\!\!\! \!\!\!\!& +~ \frac{\bm{a}^H_{jk}[n]\bm{w}_{j}[n]\bm{w}_{j}[n]^H \bm{a}_{jk}[n]}{\Vert \bm{u}_j[n] - \bm{g}_k \Vert^2 + H_j^2}.
\label{eq:sensing_SINR}
\end{eqnarray} 
On the right-hand-side of \eqref{eq:sensing_SINR}, the first and second terms are the distance-normalized beampattern gains resulted by Alice and Jack, respectively, which signify the incident signal powers arrived at target $k$.

\subsection{Problem Formulation}

\textcolor{black}{In this work, we design the dual-UAV trajectory and beamforming to maximize the ASR, while ensuring sensing performance. Specifically, the ASR maximization problems are formulated for the SC and SCS purposes, respectively, subject to the corresponding
 system constraints.}

In the SC phase, the objective is to maximize the ASR $R_{\rm{\rm{asr}}}^c=\frac{1}{N_c}\sum_{n=1}^{N_c} {R}^c[n]$ by jointly optimizing the dual-UAV trajectory $\{ \bm{u}_a[n], \bm{u}_j [n]\}$, secure communication beamformer $\{\bm{w}_{a}[n], \bm{R}_r[n]\}$, and jamming beamformer $\{ \bm{w}_{j}[n] \}$. The ASR maximization problem is formulated as
\begin{subequations}\label{eq:P1_sub}
    \begin{align}
        (\text{P}1): &~\mathop{\max}\limits_{\{\bm{w}_{m}[n]\succeq 0, \bm{u}_m[n]\}}{R}_{\rm{asr}}^c[n]\\
        \text{s.t.}&~\Vert\bm{w}_a[n]\Vert^2\le P_{a}^{\rm max},\!\!\!\!\quad\forall n \in\mathcal{N}_c\label{eq:Pa_max_constraint_c},\\
        &~\Vert\bm{w}_j[n]\Vert^2\le P_{j}^{\rm max}, \!\!\!\! \quad \forall n \in\mathcal{N}_c,\label{eq:Pj_max_constraint_c} \\ 
&~\textcolor{black}{\eqref{position_I_F},~\eqref{displacement},~\text{and}~\eqref{collision}}.
    \end{align}
\end{subequations}
In problem (P$1$), constraints \eqref{eq:Pa_max_constraint_c} and \eqref{eq:Pj_max_constraint_c} impose the transmit power limits for Alice and Jack, respectively.

In the SCS phase, the transmitted signal of Alice includes the information signal $s_a[n]$ and the sensing signal $\bm{s}_r[n]$. Consequently, the constraint $\zeta(\bm{u}_a[n], \bm{u}_j[n], \bm{g}_k) \geq \Gamma$ is enforced to ensure the sensing performance requirements, where $\Gamma$ is a required threshold for the distance-normalized sum beampattern gain. The ASR maximization problem in the SCS phase can be written as
\begin{subequations}\label{eq:P2_sub}
    \begin{align}
        &\!\!\!\!(\text{P}2): ~\mathop{\max}\limits_{\{\bm{w}_{m}[n],\bm{R}_r[n]\succeq 0, \bm{u}_m[n]\}}{R}_{\rm{asr}}^s[n]\\
        &\text{s.t.}~~\zeta(\bm{u}_a[n], \bm{u}_j[n], \bm{g}_k) 
        \ge \Gamma, ~ \forall n \in \mathcal{N}_s, ~\forall k \in \mathcal{K},\label{eq:sensing_constraint_s}\\
        &~~~~~\Vert\bm{w}_a[n]\Vert^2+ {\rm{tr}}(\bm{R}_r[n])\le P_{a}^{\rm max},~\forall n \in\mathcal{N}_s\label{eq:Pa_max_constraint_s},\\
        &~~~~~\Vert\bm{w}_j[n]\Vert^2\le P_{j}^{\rm max}, ~ \forall n \in\mathcal{N}_s,\label{eq:Pj_max_constraint_s} \\ 
             &~~~~~\textcolor{black}{\eqref{position_I_F},~\eqref{displacement},~\text{and}~\eqref{collision}}.
    \end{align}
\end{subequations}
\textcolor{black}{In \eqref{eq:P2_sub}, the ASR is $R_{\text{asr}}^s = \frac{1}{N_s}\sum_{n=1}^{N_s} R^s[n]$ in the SCS phase, with Alice's transmit power constraint \eqref{eq:Pa_max_constraint_s} differing from constraint  \eqref{eq:Pa_max_constraint_c} in the SC phase. Unfortunately, both the objective functions in problems (P$1$) and (P$2$) are non-concave due to the coupling between the dual-UAV trajectory and beamforming. Furthermore, constraint \eqref{eq:sensing_constraint_s} is also non-convex. Thus, conventional convex optimization methods are inapplicable to obtain the optimal solutions for problems (P$1$) and (P$2$).} In what follows, we propose the solution approaches for problems (P$1$) and (P$2$), respectively.  

\section{Dual-UAV Trajectory and Beamforming Optimization for The SC Purpose}
\textcolor{black}{To solve the ASR maximization problem (P$1$), which involves the joint optimization of dual-UAV trajectory and beamforming, we decompose it into three tractable subproblems: dual-UAV beamforming optimization, Alice's trajectory optimization, and Jack's trajectory optimization. After deriving the optimal solutions for each subproblem, we propose a BCD algorithm to obtain a suboptimal solution to problem (P$1$).}
  
\subsection{Optimizing Dual-UAV Beamforming}

With any given dual-UAV trajectory $\{\bm{u}_a[n],\bm{u}_j[n]\}$, problem (P$1$) reduces to the dual-UAV beamforming optimization subproblem, which can be formulated as
\begin{subequations}
\begin{align}
(\text{P}3): &~\mathop{\max}\limits_{\{\bm{w}_{a}[n],\bm{w}_{j}[n]\succeq 0\}}{R}_{\rm{asr}}^c(\bm{w}_{a}[n],\bm{w}_{j}[n]) \\		&~\text{s.t.}~\eqref{eq:Pa_max_constraint_c}~\text{and}~\eqref{eq:Pj_max_constraint_c}.
\end{align}    
\end{subequations}
\textcolor{black}{In problem (P$3$), the solutions for the dual-UAV beamformers $\{\bm{w}_a[n], \bm{w}_j[n]\}$ exhibit time slot independence, enabling per-slot optimization. For each time slot $n \in \mathcal{N}_c$, the dual-UAV beamforming subproblem can be expressed as}
\begin{subequations}
\begin{align}
(\text{P}4): &\mathop{\max}\limits_{\bm{w}_{a}[n],\bm{w}_{j}[n]\succeq 0} {R}^c(\bm{w}_{a}[n],\bm{w}_{j}[n]) &\\		
\text{s.t.}
        &~\Vert\bm{w}_a[n]\Vert^2\le P_{a}^{\rm{max}},~ \forall n \in \mathcal{N}_c,\label{eq:Pa_max_constraint2} \\
        &~\Vert\bm{w}_j[n]\Vert^2\le P_{j}^{\max},~ \forall n \in \mathcal{N}_c. \label{eq:Pj_max_constraint2}   
\end{align}    
\end{subequations}
However, due to the coupling of the dual-UAV beamformers $\bm{w}_a[n]$ and $\bm{w}_j[n]$ in the objective function of problem  (P$4$), it is impossible to directly obtain the optimal solution. In what follows, we apply SDR and SCA to tackle problem  (P$4$). 

By introducing $\bm{H}_{mq}[n]\!=\!\bm{h}_{mq}[n]\bm{h}_{mq}^H[n]$ and $\bm{W}_{m}[n]=\bm{w}_{m}[n]\bm{w}_{m}^H[n]$, where $\bm{W}_{m}[n]\succeq 0$ and ${\rm{rank}}(\bm{W}_{m}[n])= 1$, the objective function is rewritten as ${R}^c(\bm{W}_{a}[n],\bm{W}_{j}[n])$ shown in \eqref{eq:hat_R[n]_c} at the bottom of the next page. Then, problem (P$4$) can be rewritten as 
\setcounter{equation}{23}
\begin{subequations}
    \begin{align}
    (\text{P}5):&~\mathop{\max}\limits_{\bm{W}_{a}[n],\bm{W}_{j}[n]\succeq 0}\!\!\!\!\!\!{{R}}^c(\bm{W}_{a}[n],\bm{W}_{j}[n]) \\ 
    \text{s.t.} 
    &~ {\rm{tr}}(\bm{W}_{a}[n])\le P_{a}^{\max},~ \forall n \in \mathcal{N}_c.\label{eq:Pa_trace_constraint} \\
    &~ {\rm{tr}}(\bm{W}_{j}[n])\le P_{j}^{\max},~ \forall n \in \mathcal{N}_c.  \label{eq:Pj_trace_constraint} \\
    &~ {\rm{rank}}(\bm{W}_{m}[n])= 1. \label{eq:rank_constraint}
    \end{align}
\end{subequations}
\textcolor{black}{To handle the rank-one constraint \eqref{eq:rank_constraint} and non-concave objective in problem (P$5$), we employ SCA to solve it in what follows. Specifically, in iteration $t_1$, local points $\bm{W}_{a}^{(t_1)}[n]$ and $\bm{W}_{j}^{(t_1)}[n]$ are used to construct a first-order Taylor (FoT) expansion of $R^c(\bm{W}_{a}[n],\bm{W}_{j}[n])$, providing a lower-bound secrecy rate ${{R}}^c(\bm{W}_{a}[n],\bm{W}_{j}[n]) \ge \bar {{R}}^{c,(t_1)}(\bm{W}_{a}[n],\bm{W}_{j}[n])$ given by}
\begin{eqnarray} 
&{\bar{R}^{c,(t_1)}}\!\!\!\!\!\!&(\bm{W}_{a}[n],\bm{W}_{j}[n])\nonumber\\ 
&\!\!\!\!=\!\!\!\!&\!\! \!\!\!\!\log_2\Big({\rm{tr}}(\bm{H}_{ab}[n]\bm{W}_a[n]) 
   + \varphi_{jb}{\rm{tr}}(\bm{H}_{jb}[n]\bm{W}_j[n]) 
    \nonumber\\ 
&&\!\!\!\!\!\! +~ \sigma_b^2 \Big)+ \log_2\Big(\varphi_{je}{\rm{tr}}(\bm{H}_{je}[n]\bm{W}_j[n]) 
   + \sigma_e^2 \Big) 
   - a^{(t_1)} \nonumber\\ 
&&\!\!\!\!\!\! -~ {\rm{tr}}\Big(b^{(t_1)}\bm{H}_{ae}[n]\big(\bm{W}_a[n] 
   - \bm{W}_a^{(t_1)}[n]\big)\Big) \nonumber\\ 
&&\!\!\!\!\!\! -~ {\rm{tr}}\Big(\big(b^{(t_1)}[n]\varphi_{je}\bm{H}_{je}[n] 
   + c^{(t_1)}[n]\varphi_{jb}\bm{H}_{jb}[n]\big) \nonumber\\ 
&&\!\!\!\! \!\!\times~ \big(\bm{W}_j[n] 
   - \bm{W}_j^{(t_1)}[n]\big)\Big),
\label{eq:Rc_bound}
\end{eqnarray}
where
\begin{eqnarray} 			
{a}^{(t_1)} \!\!&\!\!\!\!\!=\!\!\!\!\!&\! \!\log_2 \!\Big(\!{\rm{tr}}\big(\bm{H}_{ae}[n]\bm{W}_{a}^{(t_1)}[n]\big) 
                     + \varphi_{je}{\rm{tr}}\big(\bm{H}_{je}[n]\bm{W}_j^{(t_1)}[n]\big) \nonumber\\
  &&\!\! + ~\sigma_e^2 \Big) 
     + \log_2 \!\Big(\!\varphi_{jb}{\rm{tr}}\big(\bm{H}_{jb}[n]\bm{W}_j^{(t_1)}[n]\big) 
                     + \sigma_b^2 \Big),\!\! 
\label{a^{(t_1)}}
\end{eqnarray}
  \begin{eqnarray} 
		{b}^{(t_1)}\!\!&\!\!\!\!\!\!=\!\!\!\!\!\!&\!\!\frac{\log_2(e)}{{\rm{tr}}(\bm{H}_{ae}[n]\bm{W}_{a}^{(t_1)}[n])
   +\varphi_{je}{\rm{tr}}(\bm{H}_{je}[n]\bm{W}_j^{(t_1)}[n]) +\sigma_e^2},\!\!\nonumber \\
     \label{{b}^{(t_1)}}
		\end{eqnarray}
and
  \begin{eqnarray} 
			{c}^{(t_1)}&\!\!\!\!\!\!=\!\!\!\!\!\!&\frac{\log_2(e)}{\varphi_{jb}{\rm{tr}}(\bm{H}_{jb}[n]\bm{W}_j^{(t_1)}[n])
    +\sigma_b^2}.
     \label{{c}^{(t_1)}}
		\end{eqnarray}
By replacing the objective function ${{R}}^c(\bm{W}_{a}[n],\bm{W}_{j}[n])$ with its lower-bound $\bar{{R}}^{c,(t_1)}(\bm{W}_{a}[n],\bm{W}_{j}[n])$, in iteration $t_1$ of SCA, problem (P$5$) can be approximated as 
\begin{subequations}
\begin{align}
&(\text{P}5.t_1): \max_{\bm{W}_{a}[n],\bm{W}_{j}[n]\succeq 0} 
\bar{R}^{c,(t_1)}(\bm{W}_{a}[n],\bm{W}_{j}[n]) 
\label{eq:P5_n_i_beamforming_opt} \\
&\hspace{5em} \text{s.t. } 
\eqref{eq:Pa_trace_constraint},\, \eqref{eq:Pj_trace_constraint},\, 
\text{and } \eqref{eq:rank_constraint}.
\label{eq:P5_n_i_beamforming_constraints}
\end{align}
\end{subequations}
\begin{figure*}[!b]
\hrulefill
		\normalsize
          \setcounter{equation}{22}
        \begin{eqnarray}
{R}^c(\bm{W}_{a}[n],\bm{W}_{j}[n])=\log_2\bigg(1+\frac{{\rm{tr}}(\bm{H}_{ab}[n]\bm{W}_{a}[n])}{\varphi_{jb}{\rm{tr}}(\bm{H}_{jb}[n]\bm{W}_{j}[n])+\sigma_b^2}\bigg)-
\log_2\bigg
(1+\frac{{\rm{tr}}(\bm{H}_{ae}[n]\bm{W}_{a}[n])}{\varphi_{je}{\rm{tr}}(\bm{H}_{je}[n]\bm{W}_{j}[n])+\sigma_e^2}\bigg)
\label{eq:hat_R[n]_c}	
\end{eqnarray}
\setcounter{equation}{31}
\begin{eqnarray}
\hat{R}^c(\bm{u}_a[n]) \!\!&\!\!\!\!\!\!\!\!=\!\!\!\!\!\!\!\!&\!\! \log_2\bigg(\frac{{\rm{tr}}(\bm{W}_a[n]\bm{A}_{ab}[n])}{d_{ab}^2[n]} 
                    + \frac{\varphi_{jb}{\rm{tr}}(\bm{W}_j[n]\bm{A}_{jb}[n])}{d_{jb}^2[n]}
                    + \frac{\sigma_b^2}{\beta} \bigg) 
             + \log_2\bigg( \frac{\varphi_{je}{\rm{tr}}(\bm{W}_j[n]\bm{A}_{je}[n])}{d_{je}^2[n]} 
                    + \frac{\sigma_e^2}{\beta} \bigg) \nonumber \\
&& - \log_2\bigg( \frac{{\rm{tr}}(\bm{W}_a[n]\bm{A}_{ae}[n])}{d_{ae}^2[n]}
                    + \frac{\varphi_{je}{\rm{tr}}(\bm{W}_j[n]\bm{A}_{je}[n])}{d_{je}^2[n]}
                    + \frac{\sigma_e^2}{\beta} \bigg) 
             - \log_2\bigg( \frac{\varphi_{jb}{\rm{tr}}(\bm{W}_j[n]\bm{A}_{jb}[n])}{d_{jb}^2[n]}
                    + \frac{\sigma_b^2}{\beta} \bigg)
\label{eq:R[n]_trajectory_a}
\end{eqnarray}
\end{figure*}To handle the rank-one constraint \eqref{eq:rank_constraint} in problem (P$5$.$t_1$), we propose to add a penalty term $\frac{1}{\iota _1}\sum_{m\in\{a,j\}}\big(\|\bm{W}_m\|_*+\hat{\bm{W}}_m^{(t_1)}\big)$ into the 
objective function, where 
$\iota_1$ is a control factor and $\hat{\bm{W}}_m^{(t_1)} = \Vert\bm{W}_m\Vert_2 - {\rm{tr}}\big({\bm{p}^{{(t_1)}}_{{{\rm{max}},m}}(\bm{p}^{{(t_1)}}_{{{\rm{max}},m}})^H(\bm{W}_m-\bm{W}_m^{(t_1)})}\big
)$ is the FoT expansion at point $\bm{W}^{(t_1)}_m$ \textcolor{black}{\cite{Beamforming_RIS_Assiste_RSMA,Active_RIS_MISO-NOMA,Covert_communication_Full-duplex}}. In particular, $\bm{p}^{{(t_1)}}_{{{\rm{max}},m}}$ is the eigenvector corresponding to the largest eigenvalue of $\bm{W}^{(t_1)}_m$ and we adopt $\hat{\bm{W}}_m^{(t_1)}$ as an upper bound on $-\|\bm{W}_m\|_2$. When $\iota_1\rightarrow 0$, we have the fact $\|\bm{W}_m\|_*-\|\bm{W}_m\|_2=0$ to satisfy the rank-one constraint ${\rm{rank}}(\bm{W}_{m}[n]) = 1$. Consequently, the approximated optimization problem is formulated as
\setcounter{equation}{29}
\begin{subequations}
    \begin{align}
(\text{P}6.t_1):~&\mathop{\max}\limits_{\bm{W}_{a}[n],\bm{W}_{j}[n]\succeq 0}{\bar{{R}}^{c,(t_1)}(\bm{W}_{a}[n],\bm{W}_{j}[n])} \nonumber\\
     &\qquad\qquad\qquad-\frac{1}{\iota_1}\sum_{m\in\{a,j\}}\big(\Vert \bm{W}_m \Vert_*+\hat{\bm{W}}_m^{(t_1)}\big). \label{eq:P6_n_i_beamforming_opt}\\
     &\qquad\qquad\text{s.t.}~\eqref{eq:Pa_trace_constraint} \text{ and } \eqref{eq:Pj_trace_constraint}.
    \end{align}
\end{subequations}
Now, problem (P$6$.$t_1$) is convex and we can apply a convex solver to obtain the optimal solution for it.

\subsection{Optimizing Alice's Trajectory}

For any given beamformers $\{\bm{W}_a[n], \bm{W}_j[n]\}$ and Jack's trajectory $\{\bm{u}_j[n]\}$, problem (P$1$) is simplified to optimize Alice's trajectory, with the corresponding subproblem formulated as
\begin{subequations}
    \begin{align}
       \!\!\!\! \!\!(\text{P}7):	~&\!\!\!\!\mathop{\max}\limits_{\{{\bm{u}_a[n]}\}}\frac{1}{N_c}\sum_{n=1}^{N_c}{R}^c(\bm{u}_a[n])
        \label{eq:30a}\\ 
 &\text{s.t.}
~\textcolor{black}{\eqref{position_I_F},~\eqref{displacement},~\text{and}~\eqref{collision}}.
    \end{align}
\end{subequations}
\textcolor{black}{However, the complicated coupling between Alice's trajectory and the steering vector in the non-concave objective function prohibits solving problem (P$7$). The objective function is first transformed into a tractable form, and then a trust-region SCA approach is used to optimize Alice's trajectory.}

To decouple the Alice's trajectory and steering vector in the objective function of problem (P$7$), the secrecy rate is first rewritten as $\hat{{R}}^c(\bm{u}_a[n]) $ in \eqref{eq:R[n]_trajectory_a} illustrated at the bottom of this page, where  $\bm{A}_{mq}[n]= {\bm{a}}_{mq}[n] {\bm{a}}^H_{mq}[n]$. \textcolor{black}{Nevertheless, $u_a[n]$ and $a_{aq}[n]$ are still coupled in $A_{aq}[n]$, which makes the secrecy rate a non-concave function.} To formulate a tractable objective function, we further rewrite the secrecy rate as
\setcounter{equation}{32}
\begin{align}
\hat{R}^c(\bm{u}_a[n]) &= \log_2(\zeta_{1b}[n]) + \log_2(\zeta_{2e}[n]) \nonumber \\
&\quad - \log_2(\zeta_{1e}[n]) - \log_2(\zeta_{2b}[n]),
\label{eq:R[n]_trajectory_alice}
\end{align}     
where
\begin{eqnarray} 
\zeta_{1q}[n] = \eta_{aq}[n]+\zeta_{2q}[n]
        \label{eq:u_a_zeta}
\end{eqnarray}
and 
\begin{eqnarray}
 \zeta_{2q}[n] =\bigg(\frac{\varphi_{jq}{\rm{tr}}\big(\bm{W}_j[n]\bm{A}_{jq}[n]\big)}{{d}^2_{jq}[n]}+\frac{\sigma_q^2}{\beta}\bigg) d^2_{aq}[n] \label{eq:u_a_tau}   
\end{eqnarray}
with
\begin{eqnarray}
\eta_{mq}[n]\!\!&\!\!\!\!\!=\!\!\!\!\!&\!\!\sum_{y=1}^M\big[\bm{W}_m[n]\big]_{y,y}+2\sum_{x=1}^{M}\sum_{y=x+1}^{M}\big\vert\big[\bm{W}_m[n]\big]_{x,y}\big\vert \nonumber\\
		&&\times \cos\bigg(\theta^{\bm
  {W}_m}_{x,y}[n]+2\pi\frac{d_m}{\lambda}\frac{H_m(y-x)}{d_{mq}[n]}\bigg).
  \label{eq:eta_q_Wa[n]}
	\end{eqnarray}
Then, the secrecy rate is approximated by using the FoT expansion of $ \hat{R}^c(\bm{u}_a[n])$ at local point $\bm{u}_a^{(t_2)}[n]$. Specifically, in iteration $t_2$ of the trust-region SCA, the approximated secrecy rate is given by 
\begin{eqnarray}
		{\hat{R}}^c(\bm{u}_a[n]) 
		&\!\!\!\! \ge \!\!\!\!&  \alpha_a^{(t_2)}[n]+\big({\bm{\rho}}^{(t_2)}_a[n]\big)^H\big(\bm{u}_a[n]-\bm{u}_a^{(t_2)}[n]\big) ~~~~ \nonumber\\
		&\!\!\!\! \triangleq \!\!\!\!& {\tilde{R}}^{c,(t_2)} (\bm{u}_a[n]), \label{eq:u_a_R[n]_ft_ua}
	\end{eqnarray}
where 
  \begin{eqnarray} 	\alpha_a^{(t_2)}\!\!&\!\!=\!\!&\!\!\log_2\big(\zeta_{1b}^{(t_2)}[n]\big)+\log_2\big(\zeta^{(t_2)}_{2e}[n]\big)\nonumber\\
  &&\!\!-\log_2\big(\zeta^{(t_2)}_{1e}[n]\big)-\log_2\big(\zeta^{(t_2)}_{2b}[n]\big)
   \label{eq:R[n]_a_alpha}
\end{eqnarray}
and
\begin{eqnarray} 
\bm{\rho}_a^{(t_2)} \!\!\!&\!\!\!= \!&\!\!\!\! \log_2(e)\Bigg(\!\!\bigg(\frac{\bm{\gamma}_{ab}^{(t_2)}[n]}{\zeta_{1b}^{(t_2)}[n]} \!-\! \frac{\bm{\gamma}_{ae}^{(t_2)}[n]}{\zeta_{1e}^{(t_2)}[n]}\bigg) \!+\! \bigg(\frac{1}{\zeta_{1b}^{(t_2)}[n]} \!-\! \frac{1}{\zeta_{2b}^{(t_2)}[n]} \bigg) \nonumber\\ 
&&\!\!\!\times \Big(\frac{\varphi_{jb}{\rm{tr}}(\bm{W}_j[n]\bm{A}_{jb}[n])}{d^2_{jb}[n]} \!+\! \frac{\sigma_b^2}{\beta}\Big)(\bm{u}_a^{(t_2)}[n] \!-\! \bm{v}_b) \nonumber \\
&&\!\!\!- \bigg(\frac{1}{\zeta_{1e}^{(t_2)}[n]} \!-\! \frac{1}{\zeta_{2e}^{(t_2)}[n]} \bigg)\Big(\frac{\varphi_{je}{\rm{tr}}(\bm{W}_j[n]\bm{A}_{je}[n])}{d^2_{je}[n]} \!+\! \frac{\sigma_e^2}{\beta}\Big) \nonumber \\ 
&&\!\!\!\times (\bm{u}_a^{(t_2)}[n] \!-\! \bm{v}_e)\!\!\Bigg),
\label{eq:R[n]_a_rho}
\end{eqnarray}
with
\begin{eqnarray}
\bm{\gamma}_{mq}^{(t_2)}[n]		\!\!&\!\!\!\!\!\!\!\!\!\!\!\!\!\!\!\! =\!\!\!\!\!\!\!\! \!\!\!\!\!\!\!\!& \!\!\frac{4\pi d_m H_m}{\lambda}\sum_{x=1}^{M}\sum_{y=x+1}^{M}\vert[\bm{W}_m[n]]_{x, y}\vert \sin\bigg(\theta^{\bm{W}_m}_{x,y}[n] \nonumber\\
		&&\! + \frac{2\pi d_m}{\lambda}\frac{H_m (y - x)}{ d_{mq}^{(t_2)}[n]}\bigg)\frac{ (y - x)(\bm{u}^{(t_2)}_m[n]-\bm{v}_q)}{(d_{mq}^{(t_2)}[n])^3}\!. ~~~~
  \label{eq:u_a_gamma}
	\end{eqnarray}
\textcolor{black}{In addition, the convexity of $\|\bm{u}_a[n] - \bm{u}_j[n]\|^2$ renders constraint \eqref{collision} non-convex. By applying the FoT expansion, a lower bound is derived as follows:
\begin{eqnarray}
  \|\bm{u}_a[n] - \bm{u}_j[n]\|^2 + (H_a-H_j)^2 \geq d_{aj}^{(t_2)}[n], ~\forall n\in \mathcal{N},  
\end{eqnarray}
where \
\begin{eqnarray}
 d_{aj}^{(t_2)}[n] &\!\!\!=\!\!\!& \|\bm{u}^{(t_2)}_a[n] - \bm{u}_j[n]\|^2 + 2(\bm{u}^{(t_2)}_a[n] - \bm{u}_j[n])^T\nonumber\\&&\times (\bm{u}_a[n] - \bm{u}^{(t_2)}_a[n]) +(H_a-H_j)^2\nonumber\\&&\ge d_{\min}^2,  ~\forall n\in \mathcal{N}. \label{collision_ua}   
\end{eqnarray}} 
~ To ensure the approximation accuracy of the secrecy rate during the iterations of the trust-region SCA approach, the update of Alice's trajectory is constrained by 
\begin{eqnarray}
    \big\Vert \bm{u}_a^{(t_2)}[n]-\bm{u}_a^{({t_2}-1)}[n] \big\Vert \le \psi^{(t_2)}_a , ~\forall n\in \mathcal{N}_c. \label{eq:u_a_trust_region}
\end{eqnarray}
In \eqref{eq:u_a_trust_region}, $\psi^{(t_2)}_a$ is the radius of the trust-region in iteration $t_2$, which is updated by $\psi^{({t_2}+1)}_a = r_a{\psi^{(t_2)}_a}$, where $0<r_a<1$ is the factor that controls the convergence speed of the trust-region SCA approach. With the formulated secrecy rate approximation in \eqref{eq:u_a_R[n]_ft_ua} and the introduced trajectory trust-region constraint in \eqref{eq:u_a_trust_region}, the subproblem of optimizing Alice's trajectory can be formulated as 

\begin{subequations}\label{eq:uav_t_a_trajectory_opt}
\begin{align}
		(\text{P8.}t_2): & \mathop{\max}\limits_{\{{\bm{u}_a[n]}\}}  \frac{1}{N_c} \sum_{n=1}^{N_c}{\tilde{R}}^{c,(t_2)}(\bm{u}_a[n])  \\
			&\text{s.t.}~ \textcolor{black}{\eqref{position_I_F},~\eqref{displacement},~\eqref{collision_ua},
   \text{ and } 
   \eqref{eq:u_a_trust_region}}.
\end{align}
\end{subequations}  
 Now, standard convex optimization solvers can be applied to obtain the optimal solution for problem (P$8$.$t_2$). 
 By setting a sufficiently small trust-region radius, the optimal solution for problem (P$8$.$t_2$) can be arrived in convergency.  
 
\subsection{Optimizing Jack's Trajectory}

With the given beamformers $\{\bm{W}_a[n], \bm{W}_j[n]\}$ and Alice's trajectory $\{\bm{u}_a[n]\}$, the ASR maximization problem (P$1$) is reduced to 
\begin{subequations}
\begin{align}
 \!\!\!\!       \!\!(\text{P}9):	~&\!\!\!\!\mathop{\max}\limits_{\{{\bm{u}_j[n]}\}}\frac{1}{N_c}\sum_{n=1}^{N_c}{R}^c(\bm{u}_j[n])
        \label{eq:51a}\\ 
 &\text{s.t.}~
\textcolor{black}{\eqref{position_I_F},~\eqref{displacement},~\text{and}~\eqref{collision}}.
\end{align}
\end{subequations}
\textcolor{black}{Given the presence of a similar coupling between $\bm{u}_j[n]$ and $\bm{a}_j[n]$ in the objective function of problem (P$9$), the trust-region SCA approach used in subsection III.B can be applied. The approximated secrecy rate at iteration $t_3$ is expressed as } 
\begin{eqnarray}
		\breve{R}^{c,(t_3)} (\bm{u}_j[n])
		&\!\!\!\! \triangleq \!\!\!\!&  \alpha_j^{(t_3)}[n]+\big({\bm{\rho}}^{(t_3)}_j[n]\big)^H\big(\bm{u}_j[n]-\bm{u}_j^{(t_3)}[n]\big),  \nonumber\\
       &\!\!\!\!  \!\!\!\!&  \label{eq:u_a_R[n]_ft_uj}
\end{eqnarray}
where 
 \begin{eqnarray} 	\alpha_j^{(t_3)}\!\!&\!\!=\!\!&\!\!\log_2\big(\zeta_{3b}^{(t_3)}[n]\big)+\log_2\big(\zeta^{(t_3)}_{4e}[n]\big)\nonumber\\
  &&\!\!-\log_2\big(\zeta^{(t_3)}_{3e}[n]\big)-\log_2\big(\zeta^{(t_3)}_{4b}[n]\big)
   \label{eq:R[n]_j_alpha}
\end{eqnarray}
and
\begin{eqnarray}    
\bm{\rho}_j^{(t_3)}\!\!\!&\!\!\!=\!\!\!&\!\!\log_2(e)\Bigg(\!\!\bigg (\frac{1}{\zeta 
 _{3b}^{(t_3)}[n]}\!-\!\frac{ 1}{\zeta  _{4b}^{(t_3)}[n]} \bigg)\!\bigg(\varphi_{jb}\bm{\gamma}_{jb}^{(t_3)}[n]+\frac{\sigma_b^2}{\beta}\nonumber\\
 \!\!\!\!&&\!\!\times(\bm{u}_j^{(t_3)}[n]-\bm{v}_b)\!\!\bigg)\!+\!\frac{{\rm{tr}}\big(\bm{W}_a[n]\bm{A}_{ab}[n]\big)(\bm{u}_j^{(t_3)}[n]-\bm{v}_b)}{{d}^2_{ab}[n]\zeta 
 _{3b}^{(t_3)}[n]}\nonumber\\
 \!\!\!\!&&\!\!-\bigg (\frac{1}{\zeta 
 _{3e}^{(t_3)}[n]}\!-\!\frac{ 1}{\zeta  _{4e}^{(t_3)}[n]} \bigg)\!\bigg(\varphi_{je}\bm{\gamma}_{je}^{(t_3)}[n]+\frac{\sigma_e^2}{\beta}\nonumber\\
\!\!\!\!&&\!\!\times(\bm{u}_j^{(t_3)}[n]-\bm{v}_e)\!\!\bigg)\nonumber\\
 \!\!\!\!&&\!\!-\frac{{\rm{tr}}\big(\bm{W}_a[n]\bm{A}_{ae}[n]\big)(\bm{u}_j^{(t_3)}[n]\!-\!\bm{v}_e)}{{d}^2_{ae}[n]\zeta 
 _{3e}^{(t_3)}[n]}\Bigg). 
 \label{eq:R[n]_j_rho}
		\end{eqnarray}
\textcolor{black}{Then, the non-convex constraint \eqref{collision} can be reformulated as
\begin{eqnarray}
 d_{aj}^{(t_3)}[n] &\!\!\!=\!\!\!& \|\bm{u}_a[n] - \bm{u}^{(t_3)}_j[n]\|^2 - 2(\bm{u}_a[n] - \bm{u}^{(t_3)}_j[n])^T\nonumber\\&&\times (\bm{u}_j[n] - \bm{u}^{(t_3)}_j[n]) +(H_a-H_j)^2\nonumber\\&&\ge d_{\min}^2, ~\forall n\in \mathcal{N}.\label{collision_uj}   
\end{eqnarray} }    
Furthermore, the update of Jack's trajectory in iteration $t_3$ is constrained by 
\begin{eqnarray}
    \big\Vert \bm{u}_j^{(t_3)}[n]-\bm{u}_j^{(t_3-1)}[n] \big\Vert \le \psi^{(t_3)}_j , ~\forall n\in \mathcal{N}_c. \label{eq:u_j_trust_region}
\end{eqnarray}
In \eqref{eq:u_j_trust_region}, the trust-region radius is updated by $\psi^{({t_3}+1)}_j = r_j{\psi^{(t_3)}_a}$, where $0<r_j<1$ is the factor to control the convergence speed. Then, the subproblem of optimizing Jack's trajectory can be reformulated as  
\begin{subequations}\label{eq:uav_t_j_trajectory_opt}
		\begin{align}
			\!\!\!\!(\text{P10.}t_3): & \mathop{\max}\limits_{\{{\bm{u}_j[n]}\}} 
   \frac{1}{N_c}\sum_{n=1}^{N_c} {\breve{R}}^{c,(t_3)}(\bm{u}_j[n])  \\
			&\text{s.t.}~ \textcolor{black}{\eqref{position_I_F},~\eqref{displacement},~\eqref{collision_uj},
			\text{ and } \eqref{eq:u_j_trust_region}}.
		\end{align}
	\end{subequations}
Now, problem (P$10$.$t_3$) can be effectively addressed by using standard convex optimization solvers.
 
\subsection{BCD Algorithm }
\begin{algorithm}[tt]
    { \caption{BCD Optimization Algorithm for Problem (P$1$) }
    \scriptsize 
        \begin{algorithmic}[1]
            \State \textbf{Initialize}\big\{${\bm{u}}_a^{(t_0)}[n]$\big\}, \big\{${\bm{u}}_j^{(t_0)}[n]$ \big\}, and set $t_0 = 0$.
            \State\textbf{repeat} $t_0\leftarrow t_0+1$
               \State\indent\textbf{repeat}  $t_1\leftarrow t_1+1$ 
            \State\indent\quad\quad  Given  $\bm{u}^{(t_0)}_a[n]$ and $\bm{u}^{(t_0)}_j[n]$, solve problem \Statex\indent\indent\quad\quad (P$6$.$t_1$) to obtain $\bm{W}_m^{(t_1)}[n]$
            \State\indent \textbf{Until}  ${\bar{R}}^{c,(t_1+1)}-{\bar{R}}^{c,(t_1)}\le\varphi_1$
       \State\indent Update $\{{\bm{W}}_m^{(t_0)}[n]\}\!=\!\{{\bm{W}}_m^{(t_1)}[n]\}$ 
       \State\indent \textbf{repeat} $t_2\leftarrow t_2+1$
            \State\indent\quad\quad  Given $\bm{u}^{(t_0)}_j[n]$ and $\bm{W}_m^{(t_0)}[n]$, solve problem
            \Statex\indent\indent\quad\quad  (P$8$.$t_2$) to obtain $\bm{u}_a^{(t_2)}[n]$
            \State\indent\textbf{Until} $\frac{1}{N_c} \sum_{n=1}^{N_c}({\tilde{R}}^{c,(t_2+1)}[n]-{\tilde{R}}^{c,(t_2)}[n])\le\varphi_2$
        \State\indent Update $\bm{u}_{a}^{(t_0)}[n]=\bm{u}_{a}^{(t_2)}[n]$
                \State\indent\textbf{repeat} $t_3\leftarrow t_3+1$
            \State\indent\quad\quad  Given $\bm{u}^{(t_0)}_a[n]$ and $\bm{W}_m^{(t_0)}[n]$, solve problem  
            \Statex\indent\indent\quad\quad (P$10$.$t_3$) to obtain $\bm{u}_j^{(t_3)}[n]$
                   \State \indent\textbf{Until} $\frac{1}{N_c} \sum_{n=1}^{N_c}({\breve{R}}^{c,(t_3+1)}[n]-{\breve{R}}^{c,(t_3)}[n])\le\varphi_3$
        \State\indent Update $\bm{u}_{j}^{(t_0)}[n]=\bm{u}_{j}^{(t_3)}[n]$
            \State\textbf{Until} $R^{c,(t_0+1)}_{\rm{asr}}-R^{c,(t_0)}_{\rm{asr}}\le \varphi_0$
            \State\textbf{Output} final results $\bm{W}_{m}^{(t_0)}[n], \bm{u}_{m}^{(t_0)}[n], \text{and}~\bm{u}_{j}^{(t_0)}[n]$, 
            \Statex $n\in \mathcal{N}_c$.        
        \end{algorithmic}
    }
\end{algorithm}

\textcolor{black}{To obtain the suboptimal solution for problem (P$1$), a BCD algorithm is proposed. The detailed steps of this algorithm are summarized in Algorithm $1$ \cite{Beamforming_RIS_Assiste_RSMA}.} In the proposed BCD algorithm,  the dual-UAV trajectory and beamformers are iteratively updated. Specifically, problem (P$6$.$t_1$) is solved to optimize the  beamformers $\bigl\{\bm{W}^{(t_1)}_m[n]\bigr\}$ with any given dual-UAV trajectory $\bigl\{\bm{u}^{(t_0)}_a[n], \bm{u}^{(t_0)}_j[n]\bigr\}$. Then, 
problem (P$8$.$t_2$) is solved to optimize Alice's trajectory $\bigl\{\bm{u}^{(t_2)}_a[n]\bigr\}$ with any fixed $\bigl\{\bm{W}^{(t_0)}_m[n]\bigr\}$ and $\bigl\{\bm{u}^{(t_0)}_j[n]\bigr\}$. As such, problem (P$10$.$t_3$) is solved to optimize Jack's trajectory $\bigl\{\bm{u}^{(t_3)}_j[n]\bigr\}$ with any given $\bigl\{\bm{W}^{(t_0)}_m[n]\bigr\}$ and $\bigl\{\bm{u}^{(t_0)}_a[n]\bigr\}$. 

The iteration of the BCD algorithm terminates when the improvement in the ASR falls below a predefined threshold. 
\textcolor{black}{The computational complexity of the proposed BCD algorithm is analyzed systematically in what follows \textcolor{black}{\cite{complexity,complexity_Terahertz}}. For problem (P$6$.$t_1$), which involves $2M^2$ optimization variables and $2$ convex constraints, the computational complexity is $\mathcal{O}\big(\!\log(1/\varrho)N_c(2M^2+2)^{3.5}\big)$, where  $\varrho$ represents the convergence precision. \textcolor{black}{Since problems (P$8$.$t_2$) and (P$10$.$t_3$) have identical structures with $2N_c$ optimization variables and $ 2N+N_c+1 $ convex constraints, the computational complexity is  $\mathcal{O}\big(\!\log(1/\varrho)(2N+3N_c+1)^{3.5}\big)$.}}

\section{Dual-UAV Trajectory and Beamforming Optimization for The SCS Purpose}

\textcolor{black}{To achieve target sensing within the task period, the dual-UAV trajectory and beamforming must be optimized for the SCS purpose. The hybrid monostatic-bistatic radar configuration during the SCS phase ensures secure communication while enabling cooperative sensing by reusing Jack's AN signal for sensing, which enhances resource efficiency. However, solving problem (P$2$) for optimal dual-UAV trajectory and beamforming is challenging due to complicated coupling and non-convex expressions. We assume $\Delta_{N_s} = N_s / K$ time slots are allocated to detect each target in the SCS phase. Based on the dual-UAV trajectory optimized in the SC phase, suitable dual-UAV locations for SCS are selected by setting $N_c = N$. The dual-UAV beamforming optimization in problem (P$2$) can then be solved separately with given sensing locations. This section first determines the dual-UAV locations for the SCS purpose by formulating a weighted distance minimization problem and proposing a greedy algorithm to solve it. Then, the dual-UAV beamforming optimization is addressed for the SCS purpose with the determined locations.
}

\subsection{Optimizing Dual-UAV Sensing Locations}

\textcolor{black}{In the SCS phase, both the secure communication and target sensing performance must be guaranteed. This involves maximizing the ASR while ensuring that the distance-normalized sum-beampattern gain remains above a predefined threshold. Since both performance metrics depend on the distances between UAVs and ground nodes, adjusting these distances enables tuning of the performance trade-off. Specifically, shorter distances between Alice and Bob improve communication security, while shorter distances between UAVs and sensing targets enhance beampattern gain. To address this trade-off, we formulate a weighted distance minimization problem to optimize UAV sensing locations.}

\textcolor{black}{To balance the trade-off between secure communication and sensing performance, a weighted distance is introduced as}
\setcounter{equation}{51}
\begin{eqnarray}
d_k^{\rm{scs}}[n]=\tau\big(d_{ak}[n]+d_{jk}[n]\big)+(1-\tau)d_{ab}[n]
, ~~\label{eq:weighted distances}
\end{eqnarray}
where $\tau$ is a weighting factor satisfying $0 \le \tau \le 1$. \textcolor{black}{In \eqref{eq:weighted distances}, the distances $d_{ak}[n]$, $d_{jk}[n]$, and $d_{ab}[n]$ are calculated using the dual-UAV trajectory optimized for the SC purpose in Section III. Assuming $N_c = N$, equivalently, $\forall n \in {\cal N}_c$, all dual-UAV locations from the SC phase become potential sensing candidates. Thus, determining suitable sensing locations is equivalent to selecting appropriate time slots for sensing, which leads to the following weighted distance minimization problem:}\begin{subequations}\label{eq:min weighted distance}
\begin{align}
    (\text{P11}):~ & \mathop{\min}\limits_{\mathcal{N}_s}  \sum_{n\in\mathcal{N}_s}\sum_{k=1}^{K}  d_k^{\rm{scs}}[n]  \\
    & \text{s.t.}~~ |\mathcal{N}_s|=N_s. \label{eq: constrait min weighted distance}
\end{align}
\end{subequations}
\begin{figure*}[!b]
\hrulefill
\normalsize
\setcounter{equation}{54} 
 \begin{eqnarray}
    R^s(\bm{W}_{a}[n],\bm{W}_{j}[n],\bm{R}_{r}[n])\!&\!\!\!\!=\!\!\!\!&\!\log_2\!\bigg(1+\frac{{\rm{tr}}(\bm{H}_{ab}[n]\bm{W}_{a}[n])}{\varphi_{rb}{\rm{tr}}(\bm{H}_{ab}[n]\bm{R}_{r}[n])+\varphi_{jb}{\rm{tr}}(\bm{H}_{jb}[n]\bm{W}_{j}[n])+\sigma_b^2}\bigg)\nonumber\\
   &&-
\log_2\!\bigg
(1+\frac{{\rm{tr}}(\bm{H}_{ae}[n]\bm{W}_{a}[n])}{\varphi_{re}{\rm{tr}}(\bm{H}_{ae}[n]\bm{R}_{r}[n])+{\varphi_{je}\rm{tr}}(\bm{H}_{je}[n]\bm{W}_{j}[n])+\sigma_e^2}\bigg) \label{eq:hat_R[n]}	
\end{eqnarray} 
 
\begin{eqnarray}
{\bar{R}^{s,(t_4)}}(\bm{W}_{a}[n],\bm{W}_{j}[n],\bm{R}_r[n])&\!\!\!\!\!\!=\!\!\!\!\!\!&\log_2\Big({\rm{tr}}(\bm{H}_{ab}[n]\bm{W}_{a}[n])+\varphi_{jb}{\rm{tr}}(\bm{H}_{jb}[n]\bm{W}_{j}[n])+\varphi_{rb}{\rm{tr}}(\bm{H}_{ab}[n]\bm{R}_r[n]) +\sigma_b^2 \Big)\nonumber\\
&&+
\log_2\Big(\varphi_{re}{\rm{tr}}(\bm{H}_{ae}[n]\bm{R}_{r}[n])\nonumber + \varphi_{je}{\rm{tr}}(\bm{H}_{je}[n]\bm{W}_{j}[n]) +\sigma_e^2 \Big)-{a}^{(t_4)}\nonumber\\  
&&-
  {\rm{tr}}\Big({b}^{(t_4)}[n]\bm{H}_{ae}[n]\big(\bm{W}_{a}[n]-\bm{W}_{a}^{(t_4)}[n]\big)\!\Big)\nonumber\\
  &&-{\rm{tr}}\Big(\!\big({b}^{(t_4)}[n]\varphi_{re}\bm{H}_{ae}[n]+{c}^{(t_4)}\varphi_{rb}\bm{H}_{ab}[n]\big)\!\big(\bm{R}_r[n]-\bm{R}_r^{(t_4)}[n]\big)\!\Big)\nonumber\\
  &&-{\rm{tr}}\Big(\!\big({b}^{(t_4)}[n]\varphi_{je}
\bm{H}_{je}[n]+{c}^{(t_4)}[n]\varphi_{jb}\bm{H}_{jb}[n]\big)\big(\bm{W}_j[n]-\bm{W}_j^{(t_4)}[n]\big)\!\Big) 
\label{R[n]_t_4_s}
\end{eqnarray}
\vspace{-0.1in}
		\begin{eqnarray} 
{a}^{(t_4)}&\!\!\!\!=\!\!\!\!&\log_2 \!\Big({\rm{tr}}\big(\bm{H}_{ae}[n]\bm{W}_{a}^{(t_4)}[n]\big) \!+\!
   \varphi_{re}{\rm{tr}}\big(\bm{H}_{ae}[n]\bm{R}_r^{(t_4)}[n]\big)+\varphi_{je}{\rm{tr}}\big(\bm{H}_{je}[n]\bm{W}_j^{(t_4)}[n]\big)\!+\!\sigma_e^2 \Big)\nonumber\\
   &&+\log_2 \!\Big(\varphi_{rb}{\rm{tr}}\big(\bm
{H}_{ab}[n]\bm{R}_{r}^{(t_4)}[n]\big)+
    \varphi_{jb}{\rm{tr}}\big(\bm{H}_{jb}[n]\bm{W}_j^{(t_4)}[n]\big)\!+\!\sigma_b^2 \Big)
\label{{a}^{(t_4)}}		
  \end{eqnarray}
   \vspace{-0.1in}
		\begin{eqnarray}  		
{b}^{(t_4)}=\frac{\log_2(e)}{{\rm{tr}}(\bm{H}_{ae}[n]\bm{W}_{a}^{(t_4)}[n])+\varphi_{re}{\rm{tr}}(\bm{H}_{ae}[n]\bm{R}_r^{(t_4)}[n])
   +\varphi_{je}{\rm{tr}}(\bm{H}_{je}[n]\bm{W}_j^{(t_4)}[n]) +\sigma_e^2}
     \label{{b}^{(t_4)}}
		\end{eqnarray}
   \vspace{-0.1in}
   \begin{eqnarray} 
{c}^{(t_4)}=\frac{\log_2(e)}{\varphi_{rb}{\rm{tr}}(\bm{H}_{ab}[n]\bm{R}_{r}^{(t_4)}[n])+\varphi_{jb}{\rm{tr}}(\bm{H}_{jb}[n]\bm{W}_j^{(t_4)}[n])
    +\sigma_b^2}
     \label{{c}^{(t_4)}}
		\end{eqnarray}
  \end{figure*}\textcolor{black}{Problem (P$11$) is a minimum set cover problem and an NP-hard problem. Thus, a greedy algorithm is proposed to minimize the sum weighted distance $\sum_{n \in \mathcal{N}_s} \sum_{k=1}^K d_k^{\text{scs}}[n]$ for determining suitable dual-UAV sensing locations, as summarized in Algorithm $2$. The algorithm calculates distances for each target $k$ across all time slots, stored in set $\mathcal{D}_k = \{d_k^{\text{scs}}[1], d_k^{\text{scs}}[2], \cdots, d_k^{\text{scs}}[N]\}$. For target $1$, the $\Delta_{N_s}$ time slots in $\mathcal{D}_1$ that yield the smallest $\sum_{n \in \mathcal{N}_s} d_1^{\text{scs}}[n]$ are selected. Then, these time slots are excluded from $\mathcal{D}_2, \cdots, \mathcal{D}_K$. For subsequent targets $k$ ($2 \leq k \leq K$), the same selection process is applied using the updated $\mathcal{D}_k$, excluding chosen slots from the sets $\mathcal{D}_{k+1}, \cdots, \mathcal{D}_{K}$. In the SCS phase, the dual-UAV sensing locations are extracted from the optimized dual-UAV trajectory based on the selected time slots.}

\begin{algorithm}[t]
{\scriptsize 
\caption{Greedy Algorithm for Determining Dual-UAV Sensing Locations}
\begin{algorithmic}[1]
    \State Calculate $\mathcal{D}_k = \{d_k^{\rm{scs}}[1], d_k^{\rm{scs}}[2], \cdots, d_k^{\rm{scs}}[N]\}$, $\forall k \in \{1, \cdots, K\}$; Set $\Delta_{N_s}=N_s/K$ and $k = 1$
    \State \textbf{repeat} $k \leftarrow k + 1$
    \State \indent Sort $\mathcal{D}_k$ by ascending distances  
    \State \indent Select the $\Delta_{N_s}$ time slot indices corresponding
    \Statex \indent\indent to the $\Delta_{N_s}$ smallest distances
    \State \indent Use the selected $\Delta_{N_s}$ time slot indices  
    \Statex \indent\indent to construct the set $\widetilde{\mathcal{D}}_k$    
    \State \indent Exclude $\widetilde{\mathcal{D}}_k$ from $\mathcal{D}_{k+1}, \mathcal{D}_{k+2}, \cdots, \mathcal{D}_K$
    \State \textbf{Until} $k = K$
    \State Based on the time slot indices in $\widetilde{\mathcal{D}}_1, \widetilde{\mathcal{D}}_2, \cdots, \widetilde{\mathcal{D}}_K$, 
    choose the dual-UAV sensing locations from the dual-UAV trajectory optimized for the SC phase. 
\end{algorithmic}
}
\end{algorithm}
\subsection{Optimizing Dual-UAV Beamforming}

After determining the dual-UAV sensing locations, problem (P$2$) reduces to optimize the dual-UAV beamforming for the SCS purpose. The corresponding ASR maximization subproblem can be expressed as
\setcounter{equation}{53} 
\begin{subequations}
    \begin{align}
    \!\!(\text{P}12):& \!\!\mathop{\max}\limits_{\bm{W}_{a}[n],\bm{W}_{j}[n],\bm{R}_r[n]\succeq 0}\!\!\!\!\!\!{{R}}^s(\bm{W}_{a}[n],\bm{W}_{j}[n],\bm{R}_r[n]) \\ 
    \text{s.t.}~~~~ & \!\!\!\!\!\! \frac{{\rm{tr}}(\bm{A}_{ak}[n]\bm{W}_{a}[n])+{\rm{tr}}(\bm{A}_{ak}[n]\bm{R}_{r}[n])}{d_{ak}^2[n]}
    \nonumber\\
    &\!\!\!\!\!\! + \frac{{\rm{tr}}(\bm{A}_{jk}[n]\bm{W}_{j}[n])}{d_{jk}^2[n]}
    \ge \Gamma, ~\forall n \in \mathcal{N}_s,~\forall k \in \mathcal{K}, \label{eq:sensing_SINR_constraint2_s}\\
    &\!\!\!\!\!\! {\rm{tr}}(\bm{W}_{a}[n])+{\rm{tr}}(\bm{R}_r[n])\le P_{a}^{\max},~\forall n \in \mathcal{N}_s, \label{eq:Pa_trace_constraint_s} \\
    &\!\!\!\!\!\! {\rm{tr}}(\bm{W}_{j}[n])\le P_{j}^{\max},~\forall n \in \mathcal{N}_s, \label{eq:Pj_trace_constraint_s} \\
    &\!\!\!\!\!\! {\rm{rank}}(\bm{W}_{m}[n])= 1,~\forall n \in \mathcal{N}_s. \label{eq:rank_constraint_s}
    \end{align}
\end{subequations}
For problem (P$12$), we express the secrecy rate ${R}^s(\bm{W}_{a}[n],\bm{W}_{j}[n],\bm{R}_{r}[n])$ as that in \eqref{eq:hat_R[n]}, which is shown at the bottom of this page.
To tackle the non-concave objective function, we apply SCA to obtain a suboptimal solution for problem (P$12$). In particular, the FoT expansion of the secrecy rate is adopted as a lower-bound on the objective function, which is shown in \eqref{R[n]_t_4_s} with $a^{(t_4)}$, $b^{(t_4)}$, and $c^{(t_4)}$ given by \eqref{{a}^{(t_4)}}, \eqref{{b}^{(t_4)}}, and \eqref{{c}^{(t_4)}}, respectively, as illustrated at the bottom of this page.  \textcolor{black}{In addition, to address the rank-one constraint \eqref{eq:rank_constraint_s}, a penalty term $\frac{1}{\iota_2}\sum_{m\in\{a,j\}}(\|\bm{W}_m\|_*+\hat{\bm{W}}_m^{(t_4)})$ is introduced, where $\iota_2$ is a penalty parameter and $\hat{\bm{W}}_m^{(t_4)}$ represents an upper bound on $-\|\bm{W}_m\|_2$. Using the FoT expansion at $\bm{W}_m^{(t_4)}$, $\hat{\bm{W}}_m^{(t_4)}$ is computed as $\hat{\bm{W}}_m^{(t_4)} = \Vert\bm{W}_m\Vert_2 - \mathrm{tr}\big(\bm{o}^{(t_4)}_{\mathrm{max},m}(\bm{o}^{(t_4)}_{\mathrm{max},m})^H(\bm{W}_m-\bm{W}_m^{(t_4)})\big)$, where $\bm{o}^{(t_4)}_{\mathrm{max},m}$ is the eigenvector corresponding to the largest eigenvalue of $\bm{W}^{(t_4)}_m$ \textcolor{black}{\cite{Beamforming_RIS_Assiste_RSMA,Active_RIS_MISO-NOMA,Covert_communication_Full-duplex}}.} 
Then, using SCA, problem (P$12$) can be approximated as 
\setcounter{equation}{59} 
\begin{subequations}
    \begin{align}
(\text{P}13.t_4):~&\!\!\!\!\!\!\!\!\!\mathop{\max}\limits_{\bm{W}_{a}[n],\bm{W}_{j}[n],\bm{R}_r[n]\succeq 0}\!\!\!\!\!\!{\bar{{R}}^{s,(t_4)}(\bm{W}_{a}[n],\bm{W}_{j}[n],\bm{R}_r[n])} \nonumber\\
     &  \qquad\qquad\qquad -\frac{1}{\iota _2}\!\sum_{m\in\{a,j\}}\!\big(\|\bm{W}_m\|_*+\hat{\bm{W}}_m^{(t_4)}\big) \label{eq:P13_n_i_beamforming_opt_s}\\&\text{s.t.}~\eqref{eq:sensing_SINR_constraint2_s},~ \eqref{eq:Pa_trace_constraint_s},\text{ and } \eqref{eq:Pj_trace_constraint_s}.
    \end{align}
\end{subequations}
Now, standard convex optimization tools can be applied to achieve the optimal solution for problem (P$13$.$t_4$). 

\begin{table}[tt]
\centering
\color{black}
 \scriptsize  
  \renewcommand{\arraystretch}{1.2} 
  \setlength{\tabcolsep}{3pt} 
  \caption{Simulation Parameters}  
  \vspace{-0.2cm}
  \label{tab:simulation_parameters}  
  \begin{tabular}{cc|cc}  
    \hline\hline
    \bf{Parameter} & \bf{Value} & \bf{Parameter} & \bf{Value} \\  
    \hline 
 $P_a^{\max}$ & 30 dBm & $H_a$ & 120 m \\
 $P_j^{\max}$ & 25 dBm & $H_j$ & 100 m \\
$\Gamma$ & $-20$ dBm & $\Delta t$ & 0.5 s \\
$\beta$ & $-30$ dBm & $M$ & 4 \\
$\sigma^2$ & $-80$ dBm & $T$ & $10$ s \\
$\varphi_{rb}=\varphi_{jb} = \varphi$ & $-20$ dB & ${V}_{\max}$ & 20 m/s   \\
$\varphi_{re}=\varphi_{je}$ & 0 dB & $\Delta N_s$ & 2 \\
$d_{\min}$ & 20 m & $N_s$ & 8 \\
${\bm u}_a^{_{\rm I}} = {\bm u}_j^{_{\rm I}}$-case 1 & $[0, 0]^T$ m & ${\bm u}_a^{_{\rm I}} = {\bm u}_j^{_{\rm I}}$-case 2 & $[0, 0]^T$ m \\
${\bm u}_a^{_{\rm F}} = {\bm u}_j^{_{\rm F}}$-case 1 & $[100, 0]^T$ m & ${\bm u}_a^{_{\rm F}} = {\bm u}_j^{_{\rm F}}$-case 2 & $[100, 0]^T$ m \\
$\bm{v}_b$-case 1 & $[40, 60]^T$ m & $\bm{v}_b$-case 2 & $[40, 60]^T$ m \\
$\bm{v}_e$-case 1 & $[60, 60]^T$ m & $\bm{v}_e$-case 2 & $[40, -60]^T$ m \\
$K$-case 1 & $4$  & $K$-case 2 & $6$  \\
\bottomrule
\end{tabular}
\end{table}
 
\section{Simulation Results}
In this section, we present simulation results to evaluate the system performance of the proposed SCS scheme. \textcolor{black}{The simulation parameters are given in Table \ref{tab:simulation_parameters}, unless otherwise specified. For comparison, the following benchmark schemes are considered in the simulation. }
\begin{figure}[tt]
 \begin{center}
    \includegraphics[width=2.7in]{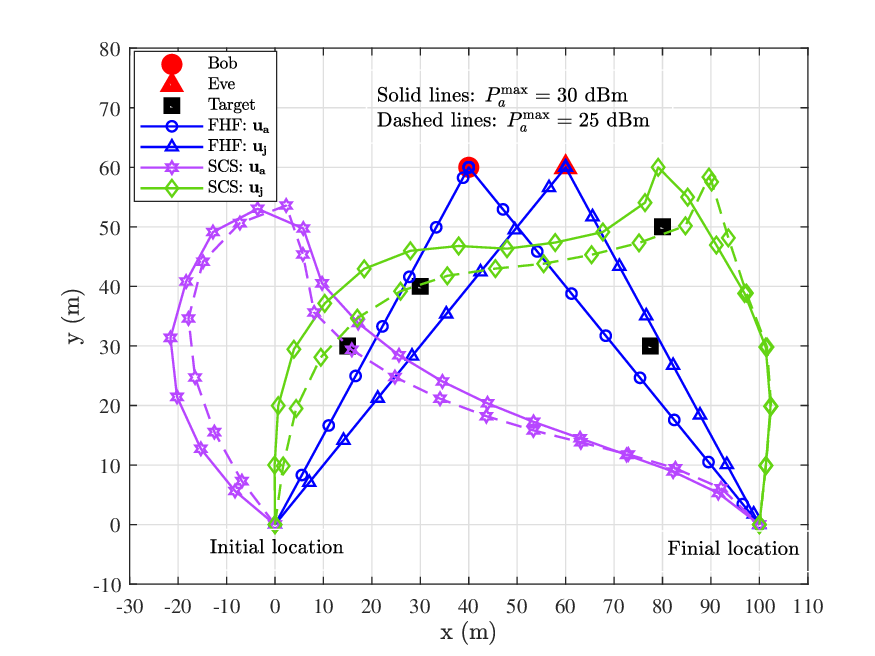}
    \caption{\textcolor{black}{UAV trajectories obtained by different schemes in case $1$.}}
    \label{fig:trajectory}
\end{center}
\vspace{-0.2in}
\end{figure}
\begin{itemize}
    \item {Fly-hover-fly (FHF): In the FHF scheme, Alice (Jack) flies from its initial location towards Bob (Eve) at the maximum speed; Then, Alice (Jack) hovers above Bob (Eve) as long as possible;  Alice (Jack) flies towards its final location at the maximum speed within the remained task period. The maximum ratio transmission (MRT) is adopted to realize dual-UAV beamforming with the main-lobes of the beams of Alice and Jack toward Bob and Eve, respectively. The dual-UAV sensing locations are determined by using the greedy algorithm.}
    \item {FHF plus beamforming: In this scheme (denoted as ``FHF+Beamforming'' in the remained figures), the dual-UAV follows the trajectory of the FHF scheme, while the dual-UAV sensing locations are obtained by using the greedy algorithm. The dual-UAV beamforming in the SC and SCS phases are obtained by solving problems (P$6$.$t_1$) and (P$13$.$t_4$), respectively.}
    \item {Single-UAV: In the single-UAV scheme, only the source UAV Alice is deployed following the FHF trajectory, while the jamming UAV Jack is omitted. The transmit beamforming of Alice is obtained by using MRT with the main-lobe of the beam toward Bob.}
\end{itemize}

\begin{figure}[tt]
 \begin{center}
    \includegraphics[width=2.7in]{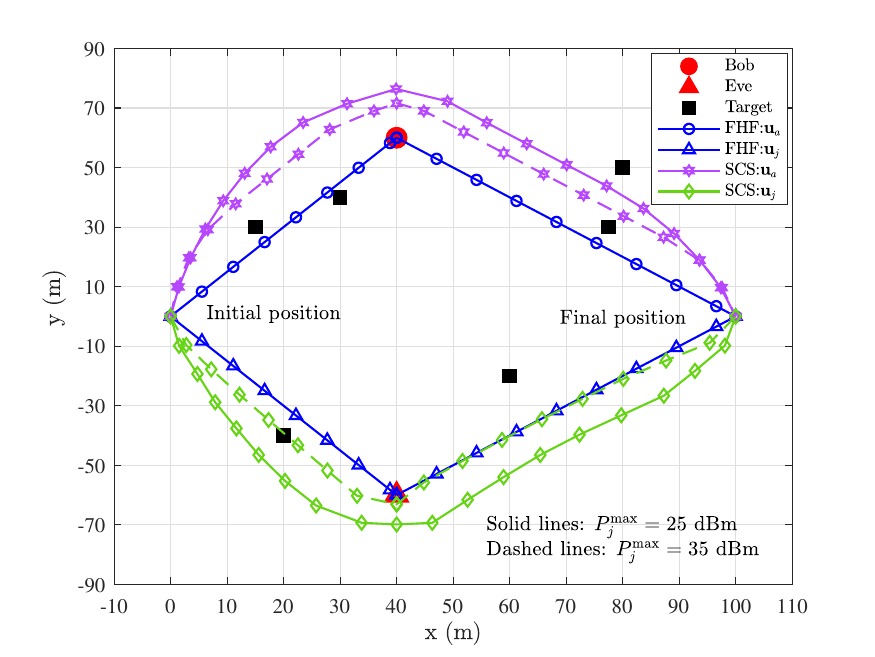}
    \caption{\textcolor{black}{UAV trajectories obtained by different schemes in case $2$.}}
    \label{fig:trajectory_case2}
\end{center}
\vspace{-0.2in}
\end{figure}

\begin{figure}[tt]
 \begin{center}
    \includegraphics[width=2.7in]{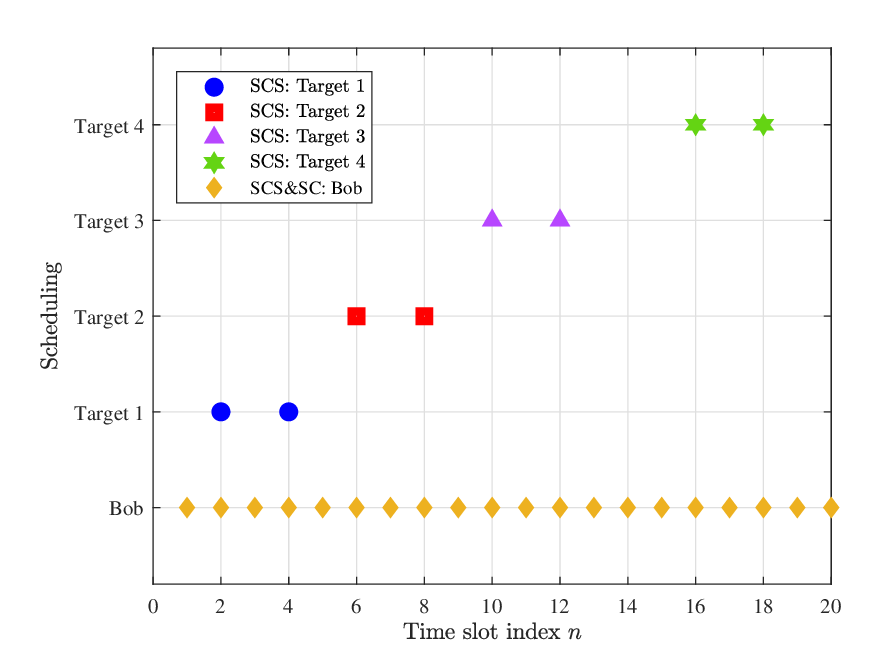}
    \caption{\textcolor{black}{Selected time slot indices for target sensing in case $1$.}}
    \label{fig:target_slot}
\end{center}
\vspace{-0.2in}
\end{figure}

\begin{figure}[tt]
 \begin{center}
    \includegraphics[width=2.7in]{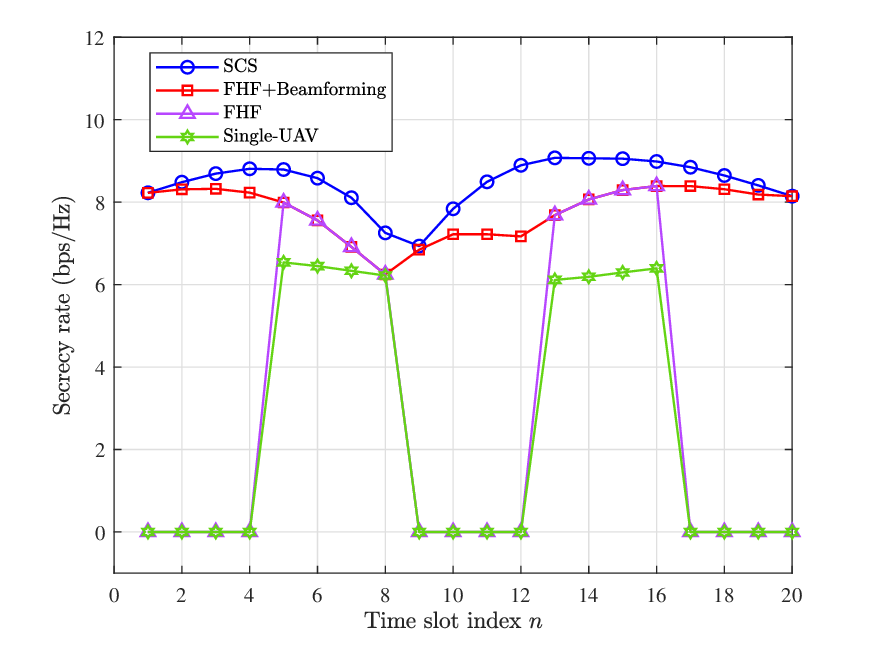}
    \caption{\textcolor{black}{Secrecy rate at each time slot in case $1$.}}
    \label{fig:time_rate}
\end{center}
\vspace{-0.2in}
\end{figure}

\begin{figure}[tt]
 \begin{center}
    \includegraphics[width=2.7in]{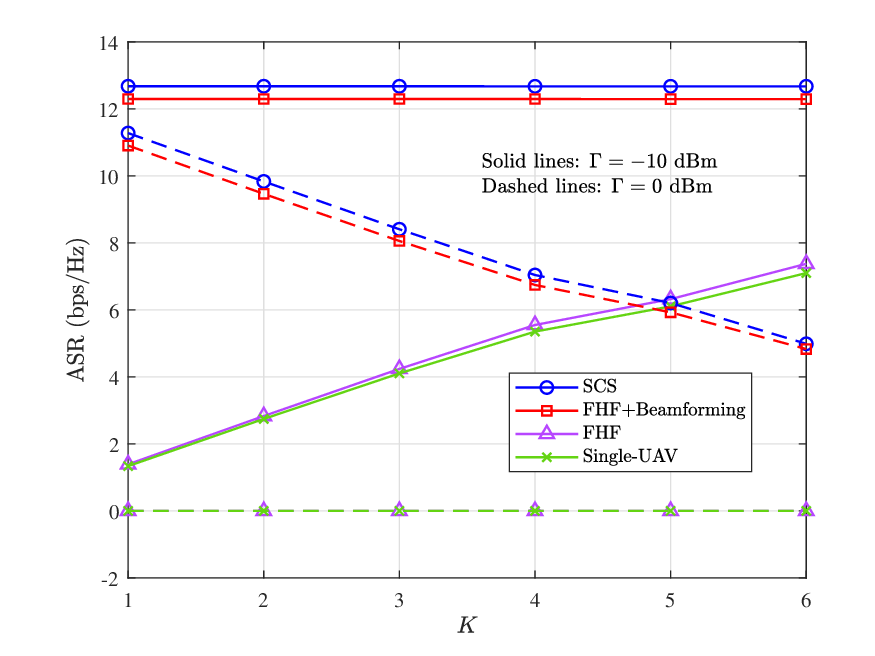}
    \caption{\textcolor{black}{ASR under the different numbers of 
ground target in case $2$.}}
    \label{fig:target_number}
\end{center}
\vspace{-0.2in}
\end{figure}

\begin{figure}[t]
 \begin{center}
    \includegraphics[width=3.2in]{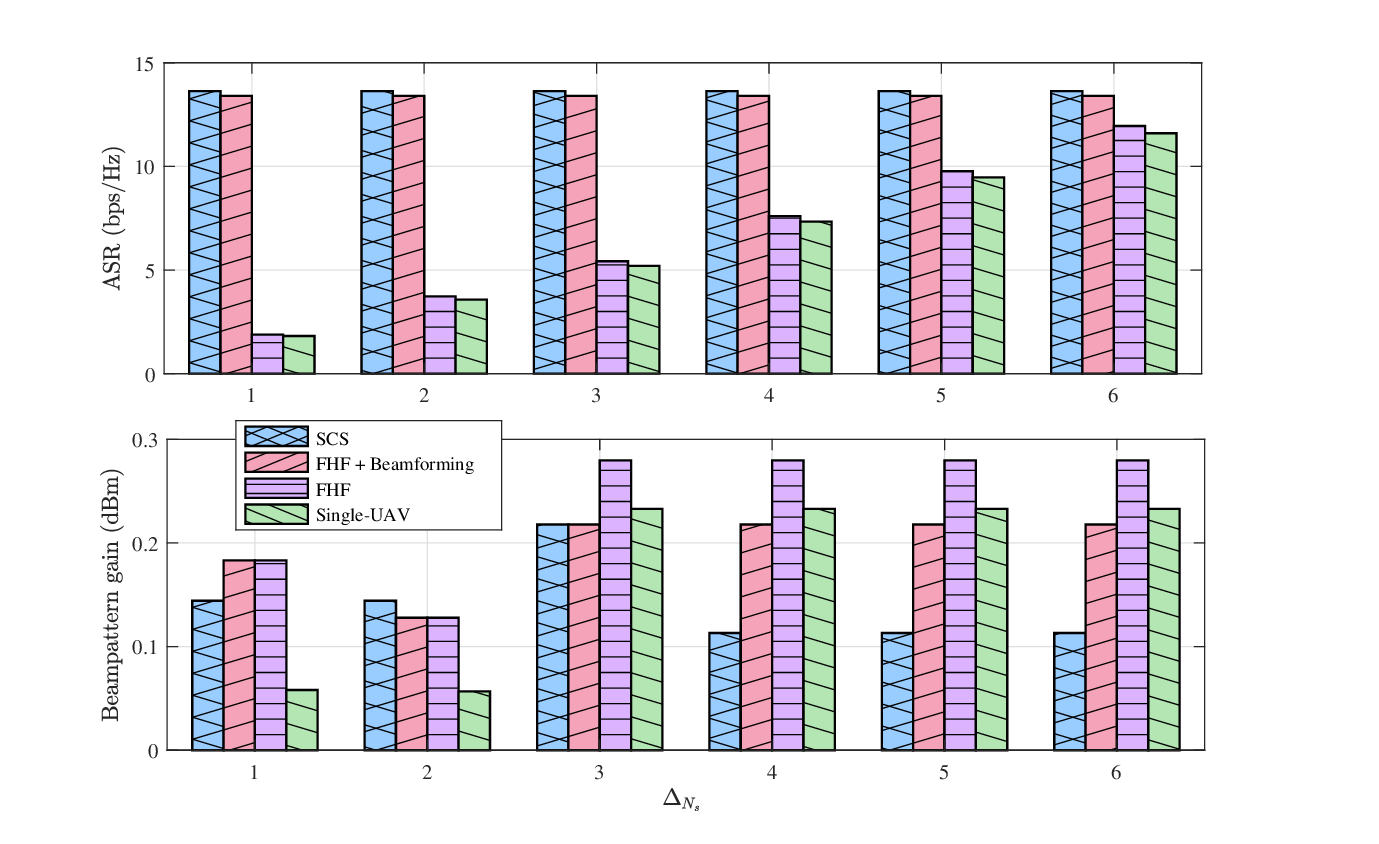}
    \caption{\textcolor{black}{ASR versus the number of time slots allocated for target sensing in case $2$.}}
    \label{fig:target_sensing_slot}
\end{center}
\vspace{-0.2in}
\end{figure}

\textcolor{black}{Fig. \ref{fig:trajectory} and \ref{fig:trajectory_case2} present the optimized dual-UAV trajectories under the proposed SCS scheme for two distinct scenarios. In Fig. \ref{fig:trajectory}, where Bob and Eve are closely located, both UAVs follow cooperative arc-shaped trajectories: Alice remains near Bob while actively evading Eve, and Jack maneuvers toward Eve to transmit AN for jamming. In Fig. \ref{fig:trajectory_case2}, where Bob and Eve are spatially separated, clearer role specialization is observed: Alice focuses exclusively on serving Bob, while Jack approaches and hovers near Eve to conduct targeted jamming, especially when sufficient jamming power is available. In both cases, the SCS scheme effectively prevents Alice and Jack from nearing Eve and Bob, respectively, unlike the FHF scheme, thereby significantly degrading the eavesdropping link quality. These results confirm that coordinated UAV mobility enables effective collaboration between Alice and Jack, enhancing overall secure communication performance.}
  
\textcolor{black}{Fig. \ref{fig:target_slot} presents the selected time slots for SCS operation, which determine the dual-UAV's sensing locations based on the pre-optimized trajectory. The results confirm that the proposed greedy algorithm effectively selects sensing time slots for each target. Meanwhile, Fig. \ref{fig:time_rate} shows that the proposed SCS scheme achieves the highest secrecy rates across all time slots, leading to the highest ASR among all benchmarks. In contrast, the FHF and single-UAV schemes yield zero secrecy rates in most slots, underscoring the necessity of dual-UAV deployment and joint trajectory and beamforming optimization for enhancing secure communication in A2G-ISAC system.}

\textcolor{black}{Fig. \ref{fig:target_number} illustrates the impact of the number of sensing targets $K$ on system performance, with $\Delta_{N_s}=3$ time slots allocated per target. Under a low beampattern gain threshold ($\Gamma = -10$ dBm), the proposed SCS scheme maintains a stable ASR as $K$ increases by allocating the minimal necessary resources, showing high efficiency. However, with a stricter threshold ($\Gamma = 0$ dBm), the ASR decreases notably as $K$ grows, due to intensified resource competition. This highlights the inherent sensing-communication trade-off in A2G-ISAC systems. Fig. \ref{fig:target_sensing_slot} further examines this trade-off in terms of time allocation. With a total duration of $T=20$ s, the ASR decreases monotonically as more slots are allocated to sensing, since communication time is compressed. Nevertheless, the SCS scheme mitigates performance degradation through joint trajectory and beamforming optimization. Notably, it consistently outperforms all benchmark schemes in ASR.}

\begin{figure}
 \begin{center}    
 \includegraphics[width=2.7in]{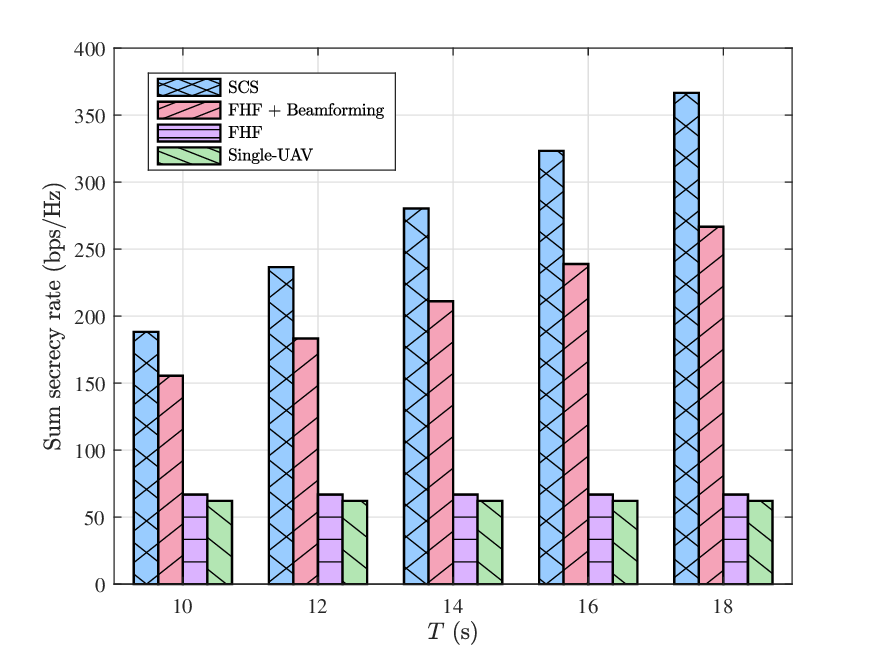}
 \caption{\textcolor{black}{Sum secrecy rate versus task period duration in case $1$.}}
 \label{fig:T}
\end{center}
\vspace{-0.2in}
\end{figure}

\textcolor{black}{Fig. \ref{fig:T} demonstrates the sum secrecy rate performance versus the task duration $T$. The results show that both the proposed SCS and FHF+Beamforming schemes achieve increasing secrecy rates as $T$ increases, with the SCS scheme consistently outperforming all benchmarks. In contrast, the FHF and single-UAV schemes maintain constant low rates across all durations, as their fixed hovering trajectories near both Bob and Eve yield nearly zero secrecy. The SCS scheme avoids this limitation by preventing proximity between two UAVs and ground nodes, thus ensuring non-zero secrecy rates throughout the mission.}
 
\begin{figure}
 \begin{center}
\includegraphics[width=2.7in]{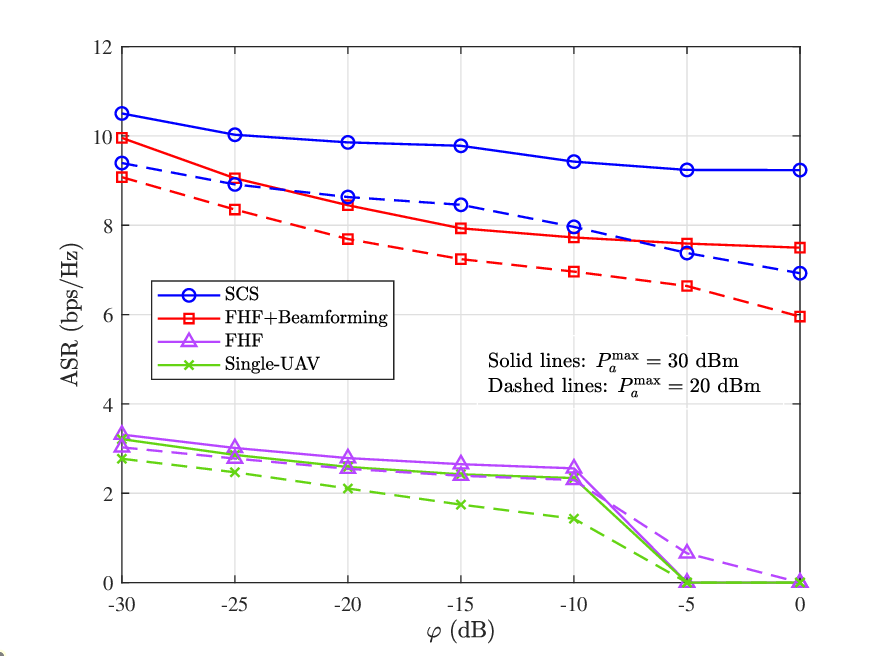}
    \caption{\textcolor{black}{ASR under different values of residual interference level in case $1$.}}
    \label{fig:cancellion factor}
\end{center}
\vspace{-0.2in}
\end{figure}

\begin{figure}
 \begin{center}
\includegraphics[width=2.7in]{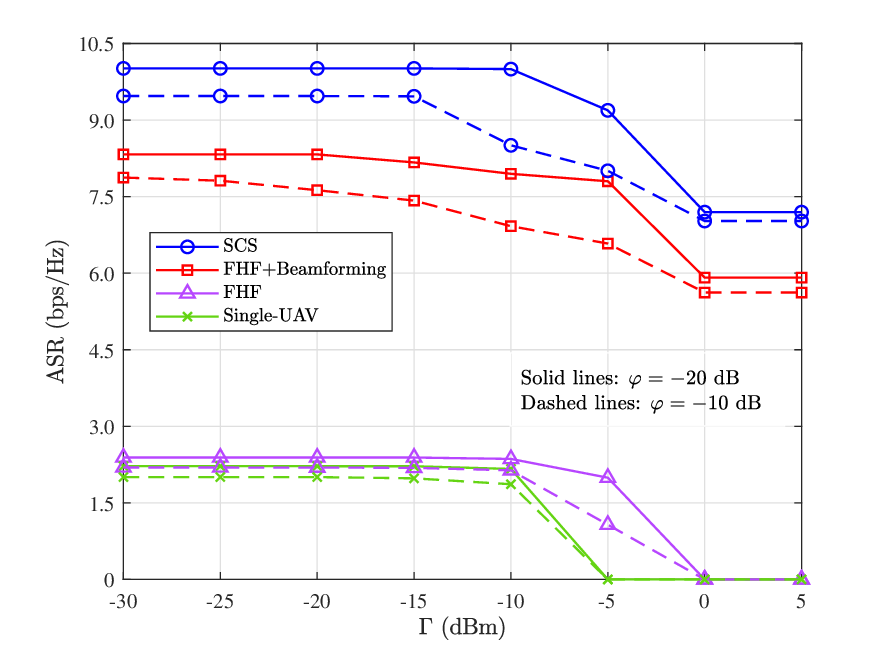}
    \caption{\textcolor{black}{Trade-off between ASR and sensing performance in case $1$.}}
    \label{fig:sensing threshold}
\end{center}
\vspace{-0.2in}
\end{figure}

\textcolor{black}{Fig. \ref{fig:cancellion factor} shows the ASR performance under different residual interference levels $\varphi$ ($\varphi_{rb}=\varphi_{jb} = \varphi$), with $T=12$ s. As $\varphi$ rises, all schemes suffer ASR degradation due to the degradation of SIC capability. The proposed SCS scheme maintains the highest ASR across all $\varphi$ values, while FHF and single-UAV schemes reduce to zero ASR at high $\varphi$. Results also indicate that lower ${P}_a^{\rm{max}}$ accelerates ASR degradation under high $\varphi$, emphasizing the importance of balancing transmit power and interference management. This validates the adaptability of the SCS scheme in complex environments through dual-UAV coordination and maneuverable jamming.}

\textcolor{black}{Fig. \ref{fig:sensing threshold} illustrates the trade-off between ASR and sensing performance at $T=14$ s. As the sum-beampattern gain threshold $\Gamma$ increases, all schemes exhibit ASR degradation under stricter sensing requirements, with higher residual interference $\varphi$ accelerating this decline. The proposed SCS scheme maintains the highest ASR across all $\Gamma$ values, though performance eventually reaches a floor under high $\Gamma$ as resources are prioritized for sensing. Conversely, lower $\Gamma$ allows more power allocation to communication, enabling ASR to approach a ceiling. These results validate the effective balance between the ASR and sensing achieved by the SCS scheme.}

\begin{figure}[t!]
	\begin{center}
		\includegraphics[width=2.7in]{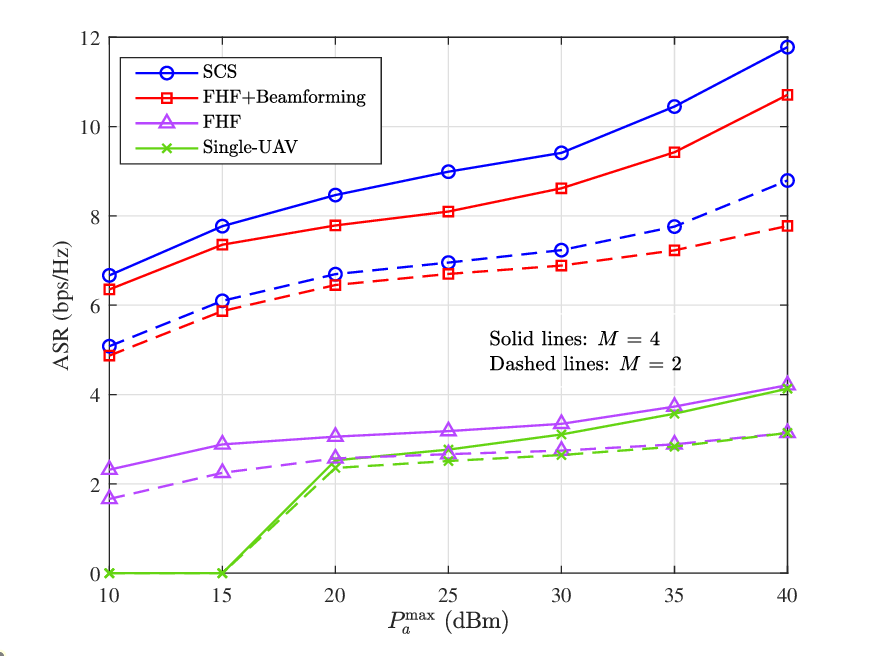}
		\caption{\textcolor{black}{ASR under the different maximum transmit powers of Alice in case $1$.}}
		\label{fig:pa}
	\end{center}
	\vspace{-0.2in}
\end{figure}

\textcolor{black}{Fig. \ref{fig:pa} examines the impact of Alice's maximum transmit power $P_a^{\max}$ on ASR. The results demonstrate that the proposed SCS scheme achieves a significantly higher ASR as $P_{a}^{\max}$ increases, with its growth rate exceeding other schemes, and using $M=4$ antennas consistently outperforms $M=2$. At low power levels, jamming provides limited improvement, and the single-UAV scheme approaches zero ASR, highlighting the necessity of multi-UAV collaboration. Through joint trajectory and beamforming optimization, the SCS scheme effectively enhances secure communication performance and power utilization efficiency.}

\begin{figure}
	\begin{center}    \includegraphics[width=2.7in]{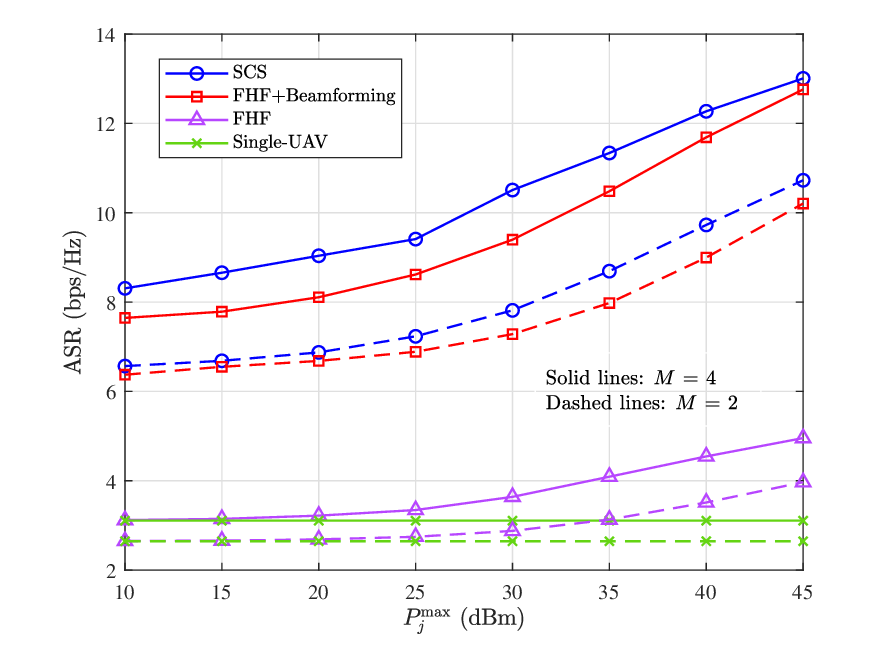}
		\caption{\textcolor{black}{ASR under the different maximum transmit powers of Jack in case $1$.}}
		\label{fig:pj}
	\end{center}
	\vspace{-0.2in}
\end{figure}

\textcolor{black}{Fig. \ref{fig:pj} shows the effect of maximum jamming power $P_j^{\max}$ on ASR. Results indicate that ASR increases with $P_j^{\max}$ for all schemes, with the proposed SCS scheme maintaining a clear advantage throughout. Using $M=4$ antennas further enhances ASR compared to $M=2$, confirming that more antennas improve eavesdropping suppression. While all schemes benefit from increased jamming power, the FHF scheme shows limited improvement due to the absence of the joint optimization of trajectory and beamforming. Furthermore, the single-UAV scheme exhibits no improvement, as it lacks both UAV cooperation and beamforming optimization.}

\begin{figure}
  	\begin{center}
  		\includegraphics[width=2.7in]{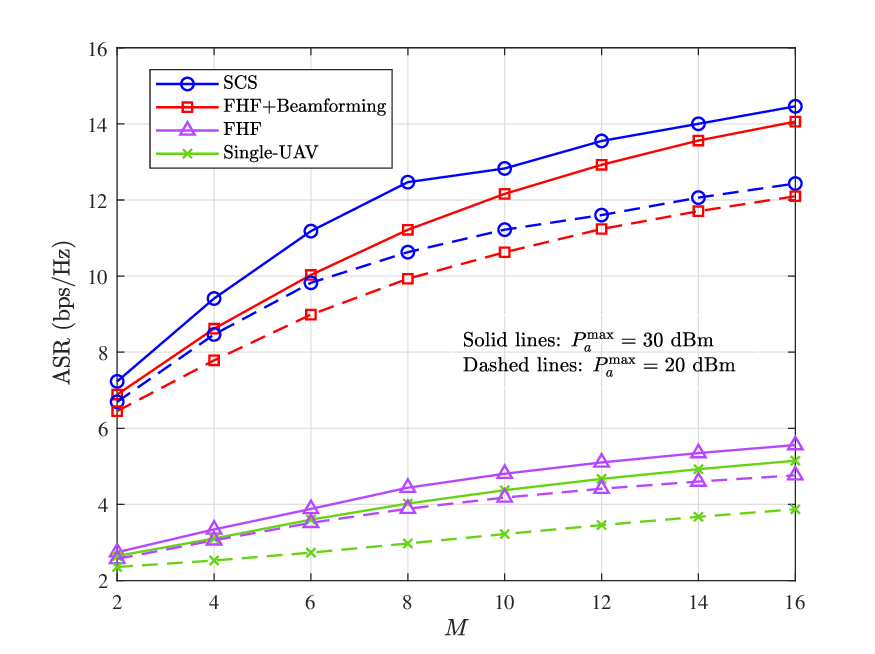}
  		\caption{\textcolor{black}{ASR under the different numbers of transmit antennas in case $1$.}}
  		\label{fig:antenna}
  	\end{center}
  	\vspace{-0.2in}
  \end{figure}

\textcolor{black}{Fig. \ref{fig:antenna} shows the ASR versus the number of transmit antennas $M$. As $M$ increases, the ASR of all schemes rises, with the proposed SCS achieving the highest ASR. Increasing $P_a^{\max}$ from $20$ dBm to $30$ dBm also boosts the ASR for all schemes. Moreover, the FHF+Beamforming scheme exhibits slower ASR growth compared to the SCS scheme, while the FHF and single-UAV schemes yield very low ASR in the considered $M$ range. These results highlight the critical role of beamforming in improving system performance. Furthermore, the simulations confirm that jointly optimizing dual-UAV trajectory and beamforming in the proposed SCS scheme significantly enhances both secure communication and target sensing performance in the A2G-ISAC system.
}

\section{Conclusions}

\textcolor{black}{In this work, we have proposed an MJ-aided SCS scheme for the A2G-ISAC system to enhance both the secure communication and target sensing performance. For the first time, the source and jamming UAVs have been configured to form a hybrid monostatic-bistatic radar for sensing multiple ground targets, while the jamming UAV is utilized to enhance secure communication performance. To balance the inherent resource allocation trade-offs in the considered A2G-ISAC system, a two-phase optimization design is adopted, thereby effectively decoupling the secure communication and sensing tasks. For the ASR maximization problem, we have sequentially addressed the optimization of the dual-UAV trajectory and dual-UAV beamforming for the SC and SCS purposes, respectively. In the SC phase, we have developed a BCD algorithm to optimize the dual-UAV trajectory and beamforming by using the SDR and trust-region SCA approaches. In the SCS phase, we have introduced a greedy algorithm to determine the dual-UAV sensing locations, along with a penalized beamforming optimization approach. Simulation results have demonstrated that the proposed SCS scheme achieves superior ASR performance compared to existing benchmarks. Furthermore, the sensing performance has been effectively enhanced by leveraging the jamming UAV with the optimized trajectory and beamforming. The simulation results have revealed the impact of the SIC residual interference levels on both the secure communication and target sensing performance. }

\begin{balance}
\bibliography{ref}

@ARTICLE{CR_NOMA_SIC,
  author={Lei, Hongjiang and Yang, Fangtao and Ansari, Imran Shafique and others},
  journal={IEEE Internet Things J.}, 
  title={Secrecy Outage Performance Analysis for Uplink {CR-NOMA} Systems With Hybrid {SIC}}, 
  year={Aug. 2023},
  volume={10},
  number={15},
  pages={13181-13195},
  doi={10.1109/JIOT.2023.3261308}}

@ARTICLE{Beamforming_RIS_Assiste_RSMA,
  author={Sun, Qiang and Liu, Hongwu and Yan, Shen and others},
  journal={IEEE Wireless Commun. Lett.}, 
  title={Joint Receive and Passive Beamforming Optimization for {RIS}-Assisted Uplink {RSMA} Systems}, 
  year={July. 2023},
  volume={12},
  number={7},
  pages={1204-1208},
  doi={10.1109/LWC.2023.3266883}}

@ARTICLE{Active_RIS_MISO-NOMA,
  author={Zhu, Miaomiao and Guo, Kefeng and Ye, Yinghui and others},
  journal={IEEE Wireless Commun. Lett.}, 
  title={Active {RIS}-Aided Covert Communications for {MISO-NOMA} Systems}, 
  year={Dec. 2023},
  volume={12},
  number={12},
  pages={2203-2207},
  doi={10.1109/LWC.2023.3314625}}

@ARTICLE{Covert_communication_Full-duplex,
  author={Kang, Xueyu and Qi, Nan and Lv, Lu and others},
  journal={IEEE Trans. Veh. Technol}, 
  title={Covert Communications in Active-{IOS} Aided Uplink {NOMA} Systems With Full-Duplex Receiver}, 
  year={May. 2025},
  volume={74},
  number={5},
  pages={8419-8424},
  doi={10.1109/TVT.2025.3528730}}

@ARTICLE{secure_UAV_ISAC_JBT,
  author={Yao, Jianping and Yang, Zeyu and Yang, Zai and others},
  journal={IEEE Wireless Commun. Lett.}, 
  title={{UAV}-Enabled Secure {ISAC} Against Dual Eavesdropping Threats: Joint Beamforming and Trajectory Design}, 
  year={Early Access, 2025},
  volume={},
  number={},
  pages={1-1},
  doi={10.1109/LWC.2025.3588758}}

@ARTICLE{complexity_Terahertz,
  author={Mamaghani, Milad Tatar and Hong, Yi},
  journal={IEEE Trans. Veh. Technol}, 
  title={Terahertz Meets Untrusted {UAV}-Relaying: Minimum Secrecy Energy Efficiency Maximization via Trajectory and Communication Co-Design}, 
  year={May. 2022},
  volume={71},
  number={5},
  pages={4991-5006},
  doi={10.1109/TVT.2022.3150011}}

@ARTICLE{add_RIS_A2G_network,
  author={Sun, Yifu and Lin, Zhi and An, Kang and others},
  journal={IEEE J. Select. Areas  Commun.}, 
  title={Multi-Functional {RIS}-Assisted Semantic Anti-Jamming Communication and Computing in Integrated Aerial-Ground Networks}, 
  year={Dec. 2024},
  volume={42},
  number={12},
  pages={3597-3617},
  doi={10.1109/JSAC.2024.3459028}}

@ARTICLE{add_secure_Ad_Hoc,
  author={Lin, Xin and Liu, Aijun and Han, Chen and others},
  journal={IEEE Trans. Wireless Commun.}, 
  title={Secure Beamforming and Anti-Jamming Coalition Formation for Air-Terrestrial Integrated Ad-Hoc Networks}, 
  year={Aug. 2025},
  volume={24},
  number={8},
  pages={6722-6736},
  doi={10.1109/TWC.2025.3555600}}

@ARTICLE{Dual_UAV_secure,
  author={Cai, Yunlong and Cui, Fangyu and Shi, Qingjiang and others},
  journal={IEEE J. Select. Areas  Commun.}, 
  title={{Dual-UAV}-Enabled Secure Communications: Joint Trajectory Design and User Scheduling}, 
  year={Sept. 2018},
  volume={36},
  number={9},
  pages={1972-1985},
  doi={10.1109/JSAC.2018.2864424}}

@ARTICLE{CS_MIMO_ISAC,
  author={Mao, Weihao and Lu, Yang and Chi, Chong-Yung and others},
  journal={IEEE Trans. Wireless Commun.}, 
  title={Communication-Sensing Region for Cell-Free Massive {MIMO} {ISAC} Systems}, 
  year={Sept. 2024},
  volume={23},
  number={9},
  pages={12396-12411},
  doi={10.1109/TWC.2024.3392330}}

@ARTICLE{Multi_UAV_Ssecure,
  author={Lei, Hongjiang and Meng, Dongyang and Ran, Haoxiang and others},
  journal={IEEE Trans. Cogn. Commun. Netw.}, 
  title={Multi-{UAV} Trajectory Design for Fair and Secure Communication}, 
  year={Early Access, 2024},
  volume={},
  number={},
  pages={1-15}}

@ARTICLE{ISAC_Advances_Challenges,
  author={Lu, Shihang and Liu, Fan and Li, Yunxin and others},
  journal={IEEE Internet Things J.}, 
  title={Integrated Sensing and Communications: Recent Advances and Ten Open Challenges}, 
  year={01 Jun. 2024},
  volume={11},
  number={11},
  pages={19094-19120},
  doi={10.1109/JIOT.2024.3361173}}

@ARTICLE{UAV_ISAC_Jamming_Attack,
  author={Mei, Hao and Zhang, Haixia and Zhou, Xiaotian and others},
  journal={IEEE Trans. Veh. Technol}, 
  title={{AoI} Minimization for Air-Ground Integrated Sensing and Communication Networks with Jamming Attack}, 
  year={Early Access, 2025},
  volume={},
  number={},
  pages={1-15},
  doi={10.1109/TVT.2025.3558061}}

@INPROCEEDINGS{UAV_ISAC_Malicious_Jamming,
  author={Liu, Yuan and Zhang, Bangning and Hu, Jingming and others},
  booktitle={Proc. 2024 IEEE 24th International Conference on Communication Technology (ICCT)}, 
  title={Beamforming Optimization for {UAV}-Enable {ISAC} Systems under Malicious Jamming Attack}, 
  year={18-20 Oct. 2024},
  address={Chengdu, China},
  number={},
  pages={1229-1233},
  doi={10.1109/ICCT62411.2024.10946330}}

@ARTICLE{ISAC_Resource_Allocation,
  author={Dong, Fuwang and Liu, Fan and Cui, Yuanhao and others},
  journal={IEEE Trans. Wireless Commun.}, 
  title={Sensing as a Service in {6G} Perceptive Networks: A Unified Framework for {ISAC} Resource Allocation}, 
  year={May. 2023},
  volume={22},
  number={5},
  pages={3522-3536},
  doi={10.1109/TWC.2022.3219463}}

@INPROCEEDINGS{UAV_secure_JTR,
  author={Yang, Wenyi and Liu, Xin and Liu, Zechen},
  booktitle={Proc. 2024 IEEE 24th International Conference on Communication Technology (ICCT)}, 
  title={Joint Trajectory and Resource Allocation Optimization for Mobile Vehicles in {UAV} Assisted Secure {ISAC} System}, 
  year={18-20 Oct. 2024},
  address={Chengdu, China},
  volume={},
  number={},
  pages={321-325},
  doi={10.1109/ICCT62411.2024.10946325}}

@ARTICLE{UAV_ISAC_Coverage_and_Security,
  author={Benaya, Ahmed M. and Hassan, Mohamed S. and Ismail, Mahmoud H. and others},
  journal={IEEE Trans. Green Commun.}, 
  title={Aerial {ISAC}: A {HAPS}-Assisted Integrated Sensing, Communications and Computing Framework for Enhanced Coverage and Security}, 
  year={Early Access, 2025},
  volume={},
  number={},
  pages={1-1},
  doi={10.1109/TGCN.2025.3551395}}

@ARTICLE{UAV_ISAC_PLS_Multiple_Eavesdroppers,
  author={Liu, Yuemin and Liu, Xin and Liu, Zechen and others},
  journal={IEEE Trans. Veh. Technol.}, 
  title={Secure Rate Maximization for {ISAC-UAV} Assisted Communication Amidst Multiple Eavesdroppers}, 
  year={Oct. 2024},
  volume={73},
  number={10},
  pages={15843-15847},
  doi={10.1109/TVT.2024.3412805}}

@ARTICLE{UAV_ISAC_Deployment_and_Precoder,
  author={Abdissa Bayessa, Gezahegn and Chai, Rong and Liang, Chengchao and others},
  journal={IEEE Internet Things J.}, 
  title={Joint {UAV} Deployment and Precoder Optimization for Multicasting and Target Sensing in {UAV}-Assisted {ISAC} Networks}, 
  year={Oct. 2024},
  volume={11},
  number={20},
  pages={33392-33405},
  doi={10.1109/JIOT.2024.3430371}}

@ARTICLE{ISAC_Dual-Functional_Wireless_Networks,
  author={Liu, Fan and Cui, Yuanhao and Masouros, Christos and others},
  journal={IEEE J.  Select. Areas  Commun.}, 
  title={Integrated Sensing and Communications: Toward Dual-Functional Wireless Networks for {6G} and Beyond}, 
  year={Jun. 2022},
  volume={40},
  number={6},
  pages={1728-1767},
  doi={10.1109/JSAC.2022.3156632}}

@ARTICLE{UAV_ISAC_IoT,
  author={Liu, Zechen and Liu, Xin and Liu, Yuemin and others},
  journal={IEEE Trans. Wireless Commun.}, 
  title={{UAV} Assisted Integrated Sensing and Communications for Internet of Things: {3D} Trajectory Optimization and Resource Allocation}, 
  year={Aug. 2024},
  volume={23},
  number={8},
  pages={8654-8667},
  doi={10.1109/TWC.2024.3352985}}

@ARTICLE{UAV_ISAC_Sky,
  author={Jing, Xiaoye and Liu, Fan and Masouros, Christos and others},
  journal={IEEE Trans. Wireless Commun.}, 
  title={{ISAC} From the {Sky}: {UAV} Trajectory Design for Joint Communication and Target Localization}, 
  year={Oct. 2024},
  volume={23},
  number={10},
  pages={12857-12872},
  doi={10.1109/TWC.2024.3396571}}

@ARTICLE{Multi_UAV_ISAC_IoTJ,
  author={Zhang, Ruizhi and Zhang, Ying and Tang, Rui and others},
  journal={ IEEE Internet Things J.}, 
  title={A Joint {UAV} Trajectory, User Association, and Beamforming Design Strategy for Multi-{UAV}-Assisted {ISAC} Systems}, 
  year={Sept. 2024},
  volume={11},
  number={18},
  pages={29360-29374}}

@INPROCEEDINGS{Multi_UAV_ISAC_GCWkshps,
  author={Ding, Weihang and Chen, Changyu and Fang, Yuan and others},
  booktitle={Proc. 2023 IEEE Globecom Workshops (GC Wkshps)}, 
  title={Multi-{UAV}-Enabled Integrated Sensing and Communications: {Joint} {UAV} Placement and Power Control}, 
  year={4-8 Dec. 2023},
  address={Kuala Lumpur, Malaysia},
  volume={},
  number={},
  pages={842-847}}

@INPROCEEDINGS{Multi_UAV_ISAC_WCSP,
  author={Ding, Weihang and Ren, Zixiang and Fang, Yuan and others},
  booktitle={Proc. 2024 16th International Conference on Wireless Communications and Signal Processing (WCSP)}, 
  title={Multi-{UAV}-Enabled Integrated Sensing and Communications: {Joint} Beamforming and {UAV} Placement Design}, 
  year={24-26 Oct. 2024},
  address={Hefei, China},
  volume={},
  number={},
  pages={770-775}}

@ARTICLE{Secure_IRS_UAV_ISAC,
  author={Zhang, Jifa and Xu, Jinlei and Lu, Weidang and others},
  journal={IEEE Trans. Wireless Commun.}, 
  title={Secure Transmission for {IRS}-Aided {UAV-ISAC} Networks}, 
  year={Sept. 2024},
  volume={23},
  number={9},
  pages={12256-12269}}

@ARTICLE{Security_ISAC_IRS_UAV,
  author={Yu, Xianglin and Xu, Jinlei and Zhao, Nan and others},
  journal={IEEE Trans. Wireless Commun.}, 
  title={Security Enhancement of {ISAC} via {IRS-UAV}}, 
  year={Oct. 2024},
  volume={23},
  number={10},
  pages={15601-15612}}

@ARTICLE{Secrecy_AAV_ISAC,
  author={Son, Chaedam and Jeong, Seongah},
  journal={IEEE Wireless Commun. Lett.}, 
  title={Secrecy Enhancement for {AAV}-Enabled Integrated Sensing and Communication Systems}, 
  year={Mar. 2025},
  volume={14},
  number={3},
  pages={696-700}}

@ARTICLE{UAVs_meet_ISAC_Secure,
  author={Wu, Jun and Yuan, Weijie and Hanzo, Lajos},
  journal={IEEE Trans. Veh. Technol.}, 
  title={When {UAVs} Meet {ISAC}: {Real}-Time Trajectory Design for Secure Communications}, 
  year={Dec. 2023},
  volume={72},
  number={12},
  pages={16766-16771}}

@ARTICLE{1,
  author={Wu, Qingqing and Zhang, Rui},
  journal={IEEE Commun. Mag.}, 
  title={Towards Smart and Reconfigurable Environment: Intelligent Reflecting Surface Aided Wireless Network}, 
  year={2020},
  volume={58},
  number={1},
  pages={106-112},
  doi={10.1109/MCOM.001.1900107}}

@ARTICLE{2,
  author={Di Renzo, Marco and Zappone, Alessio and Debbah, Merouane and Alouini, Mohamed-Slim and Yuen, Chau and de Rosny, Julien and Tretyakov, Sergei},
  journal={IEEE J. Select. Areas Commun.}, 
  title={Smart Radio Environments Empowered by Reconfigurable Intelligent Surfaces: {How} It Works, State of Research, and The Road Ahead}, 
  year={2020},
  volume={38},
  number={11},
  pages={2450-2525},
  doi={10.1109/JSAC.2020.3007211}}

@ARTICLE{4,
  author={Liu, Yuanwei and Qin, Zhijin and Elkashlan, Maged and Gao, Yue and Hanzo, Lajos},
  journal={ IEEE Trans. Veh. Technol.}, 
  title={Enhancing the Physical Layer Security of Non-Orthogonal Multiple Access in Large-Scale Networks}, 
  year={2017},
  volume={16},
  number={3},
  pages={1656-1672},
  doi={10.1109/TWC.2017.2650987}}

@ARTICLE{beamforming_design,
  author={Lyu, Zhonghao and Zhu, Guangxu and Xu, Jie},
  journal={ IEEE Trans. Veh. Technol.}, 
  title={Joint Maneuver and Beamforming Design for {UAV}-Enabled Integrated Sensing and Communication}, 
  year={Apr. 2023},
  volume={22},
  number={4},
  pages={2424-2440},
  doi={10.1109/TWC.2022.3211533}}

@ARTICLE{Hybrid_Multistatic,
  author={Temiz, Murat and Griffiths, Hugh and Ritchie, Matthew A.},
  journal={IEEE Sensors J.}, 
  title={Improved Target Localization in Multiwaveform Multiband Hybrid Multistatic Radar Networks}, 
  year={Nov. 2022},
  volume={22},
  number={21},
  pages={20785-20796},
  doi={10.1109/JSEN.2022.3206586}}

@ARTICLE{Hybrid_Bistatic_Monostatic,
  author={Thajudeen, Christopher and Hoorfar, Ahmad},
  journal={IEEE Antennas Wirel. Propag. Lett.}, 
  title={A Hybrid Bistatic–Monostatic Radar Technique for Calibration-Free Estimation of Lossy Wall Parameters}, 
  year={Nov. 2016},
  volume={16},
  number={},
  pages={1249-1252},
  doi={10.1109/LAWP.2016.2630006}}

@ARTICLE{ren2022fundamental,
  author={Ren, Zixiang and Peng, Yunfei and Song, Xianxin and others},
  journal={IEEE Trans. Wireless Commun.}, 
  title={Fundamental {CRB}-Rate Tradeoff in Multi-Antenna {ISAC} Systems With Information Multicasting and Multi-Target Sensing}, 
  year={Apr. 2024},
  volume={23},
  number={4},
  pages={3870-3885},
  doi={10.1109/TWC.2023.3312723}}

@ARTICLE{Time_slot,
  author={Chai, Rong and Cui, Xianglin and Sun, Ruijin and others},
  journal={IEEE Trans. Veh. Technol.}, 
  title={Precoding and Trajectory Design for {UAV}-Assisted Integrated Communication and Sensing Systems}, 
  year={Sept. 2024},
  volume={73},
  number={9},
  pages={13151-13163},
  doi={10.1109/TVT.2024.3390693}}

@ARTICLE{complexity,
  author={Sidiropoulos, N.D. and Davidson, T.N. and Zhi-Quan Luo},
  journal={IEEE J. Sel. Top. Signal Process.}, 
  title={Transmit beamforming for physical-layer multicasting}, 
  year={Jun. 2006},
  volume={54},
  number={6},
  pages={2239-2251},
  doi={10.1109/TSP.2006.872578}}

@ARTICLE{UAV_ISAC_EE,
  author={Liu, Yuemin and Liu, Shuai and Liu, Xin and others},
  journal={IEEE Wireless Commun. Lett.}, 
  title={Sensing Fairness-Based Energy Efficiency Optimization for {UAV} Enabled Integrated Sensing and Communication}, 
  year={Oct. 2023},
  volume={12},
  number={10},
  pages={1702-1706},
  doi={10.1109/LWC.2023.3288529}}

@ARTICLE{Jamming_multi_UAV_ISAC_secure,
  author={Wei, Zhiqiang and Liu, Fan and Liu, Chang and others},
  journal={IEEE Trans. Wireless Commun.}, 
  title={Integrated Sensing, Navigation, and Communication for Secure {UAV} Networks With a Mobile Eavesdropper}, 
  year={2024},
  volume={23},
  number={7},
  pages={7060-7078},
  doi={10.1109/TWC.2023.3337148}}

@ARTICLE{EKF_multi_UAV_JBT,
  author={Zhong, Jincheng and Wu, Jilong and Li, Yaqi and others},
  journal={IEEE Wireless Commun. Lett.}, 
  title={Joint Beamforming Design and Trajectory Optimization for {UAV}-Enabled Cell-Free {ISAC} {MIMO} Systems}, 
  year={Aug, 2025},
  volume={29},
  number={8},
  pages={1849-1853},
  doi={10.1109/LCOMM.2025.3577697}}

@INPROCEEDINGS{UAV_ISAC_Security_Aware,
  author={Yang, Zeyu},
  booktitle={Proc. 2025 5th International Conference on Sensors and Information Technology}, 
  title={Enhanced Security-Aware Beamforming and Trajectory Design for {UAV}-Assisted Integrated Sensing and Communication}, 
  year={21-23 Mar 2025},
  volume={},
  number={},
  pages={710-714},
  address= {Nanjing, China},
  doi={10.1109/ICSI64877.2025.11009397}}

@ARTICLE{LAE_ISAC_BEAM_TRAJECTORY,
  author={Cheng, Gaoyuan and Song, Xianxin and Lyu, Zhonghao and Xu, Jie},
  journal={IEEE Wireless Commun.}, 
  title={Networked {ISAC} for Low-Altitude Economy: Coordinated Transmit Beamforming and {UAV} Trajectory Design}, 
  year={Aug, 2025},
  volume={73},
  number={8},
  pages={5832-5847},
  doi={10.1109/TCOMM.2025.3541027}}
\end{balance}

\end{document}